\documentclass[twocolumn, journal]{IEEEtran}
\IEEEoverridecommandlockouts

\usepackage{color}
\usepackage{graphicx}
\usepackage{amsmath}
\usepackage{amssymb}
\usepackage{algorithm}
\usepackage{algorithmic}
\usepackage{amsmath}
\usepackage{multirow}
\usepackage{booktabs}
\usepackage{array}
\usepackage{amsthm}
\usepackage{stfloats}
\usepackage{caption}
\usepackage{bm}
\usepackage{epstopdf}
\usepackage{makecell}

\newcommand{\be}{\begin{equation}}
\newcommand{\ee}{\end{equation}}
\newcommand{\bea}{\begin{eqnarray}}
\newcommand{\eea}{\end{eqnarray}}
\newcommand{\ba}{\begin{array}}
\newcommand{\ea}{\end{array}}

\makeatletter

\newcommand{\Rmnum}[1]{\expandafter\@slowromancap\romannumeral #1@}
\makeatother

\newcommand{\RNum}[1]{\uppercase\expandafter{\romannumeral #1\relax}}

\title{Joint Beamforming Design for Intelligent Omni Surface Assisted Wireless Communication Systems\vspace{-0.7 cm}
\thanks{Part of this paper has been presented in the IEEE Global Communications Conference (GLOBECOM), 2021 \cite{Globecom}.}
\thanks{W. Cai, M. Li, and Y. Liu are with the School of Information and Communication Engineering, Dalian University of Technology, Dalian 116024, China (e-mail: wenhaocai@mail.dlut.edu.cn; mli@dlut.edu.cn; yangliu\_613@dlut.edu.cn).}
\thanks{Q. Wu is with the State Key Laboratory of Internet of Things for Smart City, University of Macau, Macau, 999078, China and Guangdong-Macau Joint Laboratory for Advanced and Intelligent Computing (email: qingqingwu@um.edu.mo).}
\thanks{Q. Liu is with the School of Computer Science and Technology, Dalian University of Technology, Dalian 116024, China (e-mail: qianliu@dlut.edu.cn).}} \author{Wenhao Cai,
        Ming Li,~\IEEEmembership{Senior Member,~IEEE,}
        Yang Liu,~\IEEEmembership{Member,~IEEE,}\\
        Qingqing Wu,~\IEEEmembership{Senior Member,~IEEE,}
        and Qian Liu,~\IEEEmembership{Member,~IEEE}}

\begin{document}
\maketitle
\thispagestyle{empty}
\begin{abstract}
Intelligent reflecting surface (IRS) has been widely considered as one of the key enabling techniques for future wireless communication networks owing to its ability of dynamically controlling the phase shift of reflected electromagnetic (EM) waves to construct a favorable propagation environment. While IRS only focuses on signal reflection, the recently emerged innovative concept of intelligent omni-surface (IOS) can provide the dual functionality of manipulating reflecting and transmitting signals. Thus, IOS is a new paradigm for achieving ubiquitous wireless communications. In this paper, we consider an IOS-assisted multi-user multi-input single-output (MU-MISO) system where the IOS utilizes its reflective and transmissive properties to enhance the MU-MISO transmission. Both power minimization and sum-rate maximization problems are solved by exploiting the second-order cone programming (SOCP), Riemannian manifold, weighted minimum mean square error (WMMSE), and block coordinate descent (BCD) methods. Simulation results verify the advancements of the IOS for wireless systems and illustrate the significant performance improvement of our proposed joint transmit beamforming, reflecting and transmitting phase-shift, and IOS energy division design algorithms.
Compared with conventional IRS, IOS can significantly extend the communication coverage, enhance the strength of received signals, and improve the quality of communication links.
\end{abstract}


\maketitle
\section{Introduction}
With the rapid development of wireless communication networks, the number of mobile devices is exponentially growing and the demand for data rate is gradually increasing.
Wireless transmission capacity has been greatly improved by several key enabling technologies, such as massive multiple-input multiple-output (MIMO), millimeter wave (mmWave) communications, and ultra-dense networks (UDN). However, the implementation of these technologies is still limited by hardware cost, energy efficiency, deployment difficulty, and computing power required for complex signal processing \cite{5G}.
Therefore, future wireless networks need fundamental innovation from the hardware level to meet the increasing capacity demands \cite{6G}.

\begin{table*}
  \centering
  \caption{A summary of representative works on IOS/STAR-RIS optimization.}\label{comp}
  \resizebox{1.65 \columnwidth}{!}{
  \begin{tabular}{|c|c|c|c|c|c|}
  \hline Reference& System setup
                  & Design objective
                  & \makecell[c]
                  {Applicable IOS\\ control mode}
                  & \makecell[c]
                  {Optimization\\ techniques}
                  & \makecell[c]
                  {$\text{Complexity}^\diamond$}\\
  \hline \cite{IOS3}, \cite{IOS1}
               & MU MISO
               & \makecell[c]{Rate maximization}
               & \makecell[c]{EED mode}
               & \makecell[c]{AO, ZF, Branch and bound (BnB)}
               & $2^M$\\
  \hline \cite{IOS2}
               & \makecell[c]{TU MISO}
               & \makecell[c]{Rate maximization}
               & \makecell[c]{EED mode}
               & \makecell[c]{BnB}
               & $2^M$\\
  \hline \cite{STAR}, \cite{STAR1}, \cite{STAR2}
               & TU MISO
               & \makecell[c]{Power minimization}
               & \makecell[c]{UED mode,\\SD mode, TD mode}
               & \makecell[c]{Penalty-based Algorithm (PA), SCA, SDR}
               & $M^{3.5}$\\
  \hline \cite{STAR3}
               & TU MIMO
               & \makecell[c]{Rate maximization}
               & \makecell[c]{UED mode,\\SD mode, TD mode}
               & \makecell[c]{Lagrange dual method, AO,\\ Penalty concave convex procedure (PCCP)}
               & -\\
  \hline \cite{STAR4}
               & MU NOMA
               & \makecell[c]{Rate maximization}
               & \makecell[c]{UED mode}
               & \makecell[c]{AO, SCA, SDR, etc.}
               & $M^{4.5}$\\
  \hline \cite{STAR5}
               & \makecell[c]{MU NOMA,\\MU OMA}
               & \makecell[c]{Coverage range}
               & \makecell[c]{UED mode}
               & \makecell[c]{One-dimensional search}
               & $M^{3}$\\
  \hline \cite{IOS4}, \cite{IOS5}
               & MU MISO
               & \makecell[c]{Secrecy energy efficiency}
               & \makecell[c]{EED mode}
               & \makecell[c]{AO, BnB, SROCR, etc.}
               & $2^M$\\
  \hline \cite{BIOS}
               & MU MISO
               & \makecell[c]{Rate maximization}
               & \makecell[c]{Bilayer-IOS\\ (Similar to EED mode)}
               & \makecell[c]{AO, WMMSE, BCD, etc.}
               & $M^2$\\
  \hline \cite{STAR6}
               & TU MIMO
               & \makecell[c]{Weighted sum secrecy rate}
               & \makecell[c]{UED mode,\\SD mode, TD mode}
               & \makecell[c]{PCCP, AO}
               & -\\
  \hline \cite{STAR7}
               & MU NOMA
               &\makecell[c]{Rate maximization}
               &\makecell[c]{UED mode}
               &\makecell[c]{Simultaneous-signal-enhancement\\-and-cancellation-based (SSECB) design}
               & -\\
  \hline \cite{STAR8}
               & TU MIMO
               &\makecell[c]{Power minimization}
               &\makecell[c]{UED mode, EED mode}
               &\makecell[c]{AO, PD, SCA}
               & $M^{3.5}$\\
  \hline Our & MU MISO
             & \makecell[c]
             {Power minimization,\\ Rate maximization}
             & \makecell[c]
             {UED mode, EED mode,\\SD mode, TD mode}
             & \makecell[c]
             {AO, WMMSE, SDR, Manifold}
             & $M^2$\\
  \hline
  \end{tabular}}
  \begin{flushleft}
  \small{$\diamond$ To fairly compare the complexity of the algorithms in different scenarios, the operations are under the assumptions $M \gg N_t, M \gg K$.}\\
  \end{flushleft}
\end{table*}

Recently, the rapid development of meta-surface introduces an innovative application in wireless communications, i.e., intelligent reflecting surfaces (IRS).
IRS is a kind of meta-surface consisting of an array of passive reflecting elements, each of which can independently induce a proper phase shift on the incident signal \cite{IRS}. Through intelligently reconfiguring the wireless propagation environment by controlling the phase shifts of reflected electromagnetic (EM) waves that impinge on the elements of surface, IRS can significantly improve the spectrum efficiency, energy efficiency, security, and reliability of wireless communication systems \cite{tutorial}. Compared with traditional relay and backscatter approaches, IRS has the advantages of easy deployment, high compatibility, environment friendly, energy-efficient, and low economic cost \cite{Pri. & Opp.}. Thus, it has attracted wide attentions in both academia and industry. In recent years, various IRS applications in practical communication scenarios have been widely studied,
including but not limited to the designs for maximizing energy efficiency \cite{IRS_ee}, spectral efficiency \cite{IRS_se}, sum-rate \cite{IRS_sr1}, \cite{IRS_sr2}, and minimizing transmit power \cite{IRS_tp1}, \cite{IRS_tp2}, etc.

\begin{figure*}[t]
  \centering
  \includegraphics[width= 5.6 in]{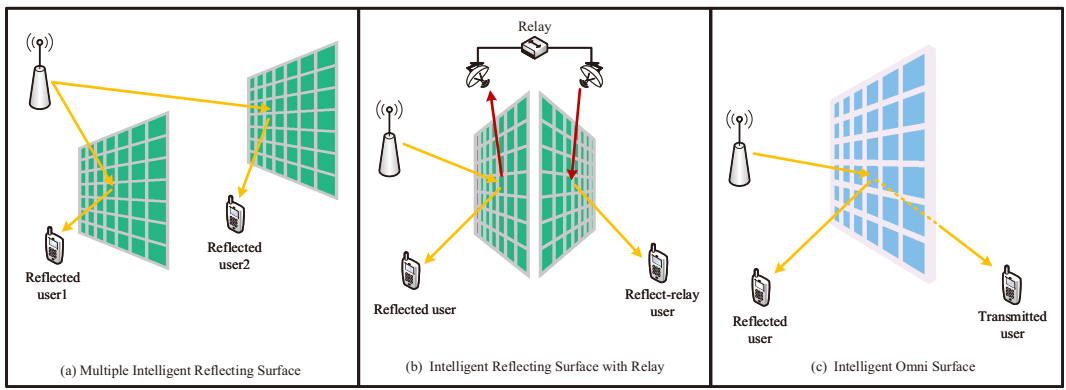}\\
  \caption{Several schemes of expending the service range of the IRS.}\label{Fig:introduction}
  \vspace{-0.4 cm}
\end{figure*}

However, in the existing studies, the function of the IRS is limited to reflect the incident signal. Hence both the source and the destination have to be located at the same side of the IRS, i.e., within the same half-space of the communication environment.
This fact limits the deployment flexibility and the service range of the IRS. Thus, researchers have proposed different schemes to expand the service range of the IRS, as shown in Fig. \ref{Fig:introduction}.
Deploying multiple IRSs \cite{multi_IRS} can provide more cooperative passive beamforming gain and a broader range of services, as shown in Fig. \ref{Fig:introduction}(a).
However, it is difficult to cooperatively control multiple IRSs in practice systems.
Relay-aided IRS architecture \cite{IRS_RELAY1} consists of two IRSs connected via a full-duplex relay, which can be deployed in a flexible way, as shown in Fig. \ref{Fig:introduction}(b).
The proposed architecture integrates the advantages of both relay and IRS, extends the coverage, and reduces the required number of IRS elements while achieving the same spectral efficiencies. However, the single-amplifier relay used in this scheme limits the number of service users.

Different from the above schemes, the intelligent omni-surface (IOS) \cite{IOS3}, which is also referred to as simultaneously transmitting and reflecting (STAR)-RIS \cite{STAR}, is a new paradigm for achieving ubiquitous wireless communications, as shown in Fig. \ref{Fig:introduction}(c).
In contrast to existing conventional IRS, IOS has the following distinct advantages.
Firstly, IOS can provide the dual-functionality of manipulating signal reflection and transmission, and the transparent substrate does not interfere aesthetically or physically with the surrounding environment \cite{STAR}, \cite{IOS_S1}.
Secondly, the coverage of IOS is extended to the entire space, thus serving both half-spaces using a single device \cite{STAR}.
More importantly, IOS can tune the phase-shifts of the reflecting/transmitting signals to generate appropriate beamforming and adjust the power ratios between reflected and transmitted signals. Thus, IOS provides new degree of freedoms (DoFs) for extending the communication coverage, enhancing the strength of the received signal, and improving the quality of service (QoS) of the communication links.

Owing to its advantages, IOS has received increasing attention recently \cite{IOS3}-\cite{STAR5}.
Two different IOS hardware implementations are proposed.
In the first IOS hardware implementation \cite{IOS3}, \cite{IOS2}, \cite{IOS1}, the reflected and transmitted energy are divided by a constant parameter, e.g., the energy is equally divided, and the ratio cannot be adjusted after the IOS deployment. The spectral efficiency maximization problems for two-user (TU) \cite{IOS2} and multi-user (MU) cases \cite{IOS1} are investigated, respectively. This assumption has better compatibility with the existing algorithms of the conventional IRS, but cannot obtain a significant performance improvement.
In the other IOS hardware implementation \cite{STAR1}-\cite{STAR5}, the reflection-to-transmission ratio for each element can be individually designed, i.e., each element has independently reflecting/transmitting amplitude and phase-shift.
The authors in \cite{STAR1} presented a general hardware implementation and two channel models for the near-field and the far-field regions.
The authors in \cite{STAR2}, \cite{STAR3} proposed three practical operating protocols for IOS and developed a heuristic algorithm in the single-user case. In addition, IOS enhanced NOMA networks \cite{STAR4}, \cite{STAR5} were investigated.
This IOS hardware implementation can achieve maximum performance improvement by introducing new DoFs in the beamforming design.

In Table \ref{comp}, we summarizes the representative works \cite{IOS3}-\cite{STAR8} on IOS/STAR-IRS optimization based on their considered system setups, design objectives, optimization techniques, etc.
However, the complex coupling between the transmit beamformers, reflecting phase-shift, transmitting phase-shift, and reflecting amplitude brings huge challenges to the algorithm design.
To the best of the authors' knowledge, there is no algorithm to solve the joint beamforming design in the IOS-assisted multi-user system with different control modes, which motivates this work.

In this paper, we propose a novel IOS-assisted communication system with multiple signal-antenna users and a multi-antenna base station (BS) where the IOS utilizes its reflective and transmissive properties to enhance the multi-user multi-input single-output (MU-MISO) downlink transmission.
We analyze the existing IOS implementations and different control modes with advantages and disadvantages. Then, we focus on the IOS implementation in which reflection/transmission amplitude and phase shift can be adjusted independently, and provide the corresponding mathematical model.
The joint BS transmit beamforming, IOS phase shift, and IOS energy division designs are investigated to improve the signal strength of users at both sides of IOS.
Our main contributions are summarized as follows:
\begin{itemize}
  \item Considering an IOS-assisted MU-MISO system, we first investigate the power minimization problem, which aims to minimize the total transmit power subject to the signal-to-interference-plus-noise ratio (SINR) constraints of all users and the IOS hardware constraints.
      A three-step algorithm is proposed to solve for the BS transmit beamforming, IOS phase-shifts, and IOS energy division by utilizing second-order cone programming (SOCP) and Riemannian manifold optimization.
  \item Next, we study the sum-rate maximization problem subject to the total transmit power constraint and IOS hardware constraints. In order to handle this non-convex NP-hard problem, we exploit the weighted minimum mean square error (WMMSE) approach to convert the original problem into a solvable multi-variable optimization, which is handled by the typical block coordinate descent (BCD) method.
      By deriving a closed-form solution for each variable, the computational complexity of our proposed algorithm is significantly reduced compared with the conventional methods.
  \item The joint design problem includes the IOS energy division and is a dual-variable optimization problem with complex coupling between variables. Thus, the widely investigated IRS beamforming design algorithms are no longer applicable, and we have developed the corresponding efficient algorithms.
      In addition, compared with previous works summarized in Table I, our proposed algorithms are compatible with different IOS control modes and significantly improve the performance of the communication system with multiple reflected/transmitted users.
  \item Finally, extensive simulation results are illustrated to exhibit the effectiveness of our proposed algorithms and validate the significant performance improvement achieved by the novel IOS implementation in the designs for the communication system with multiple reflection/transmission users.
\end{itemize}

\textit{Notation}:
Boldface lower-case and upper-case letters indicate column vectors and matrices, respectively. $\mathbb{C}$ and $\mathbb{R}^+$ denotes the set of complex and positive real numbers, respectively. $(\cdot)^*$, $(\cdot)^T$, $(\cdot)^H$, and $(\cdot)^{-1}$ denote the conjugate, transpose, conjugate-transpose operations, and the inversion of a matrix, respectively. $\mathbb{E}\{\cdot\}$ and $\mathfrak{R}\{\cdot\}$ denote statistical expectation and the real part of a complex number, respectively. $\mathbf{I}_L$ indicates an $L \times L$ identity matrix. $||\mathbf{a}||$ denotes the $\ell_2$ norm of a vector $\mathbf{a}$.
In addition, $\odot$ denotes the Hadamard product.

\section{Intelligent Omni Surface and System Model}
In this section, we first introduce the hardware implementation and the mathematical model of IOS.
Different IOS control modes are presented to show their advantages in different communication scenarios.
In addition, an IOS-assisted multi-user communication system is modeled as an example to explain how the IOS assists the communication system.

\subsection{Intelligent Omni Surface}
\begin{figure}[t]
\centering
  \includegraphics[height= 2 in]{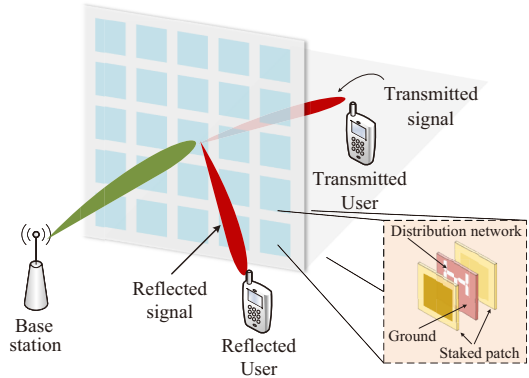}
  \caption{The schematic diagram of the IOS working principle.}\label{fig:IOS}
  \vspace{-0.3 cm}
\end{figure}
IOS is a two-dimensional meta-surface composed of a large number of controllable electromagnetic scattering elements \cite{IOS2}, as shown in Fig. \ref{fig:IOS}.
For each element of IOS, it is usually composed of radiation unit, distribution network and reference plane \cite{IOS_Hardware}. The radiation unit is usually composed of multi-layer patch radiation structures and the ground plane of multi-layer superposition is adopted.
When a signal impinges from either side of the surface, a fraction of the incident signal is reflected back propagating within the same half-space as that of the transmitter, while the other signal penetrates through the surface and continuously propagates to the opposite side of the incident signal \cite{IOS_S1, IOS_S2}. Similar to the traditional IRS, IOS can provide appropriate beamforming by cooperatively configuring the phase-shifts of elements and introducing the phase gradient on the wavelength scale.
Particularly, since IOS can simultaneously implement reflection and transmission, the distribution network can be regarded as a dual-port network.
The function of the distribution network is to split the incident signal into the reflecting part and transmitting part, and then adjust their amplitudes and phase-shifts, respectively \cite{IOS_Hardware}.
Therefore, an IOS can generate the desired reflective beamforming and transmissive beamforming by appropriately changing the reflection phase-shift $\varphi_\mathrm{r}$ and transmission phase-shift $\varphi_\mathrm{t}$ of each element.
The state-of-the-art technology can successfully realize separately controlling $\varphi_\mathrm{r}$ and $\varphi_\mathrm{t}$ \cite{IOS_Hardware2}, \cite{IOS_Hardware3}, while their amplitudes are coupled.
Thus, the reflective and transmissive coefficients $\phi_\mathrm{r}$ and $\phi_\mathrm{t}$ that describe the effect of an IOS element on the incident EM waves are given by
\begin{subequations}
\begin{align}
    \phi_\mathrm{r} &\triangleq \zeta \varphi_\mathrm{r},\\
    \phi_\mathrm{t} &\triangleq \eta  \varphi_\mathrm{t},
\end{align}
\end{subequations}
where $|\varphi_\mathrm{r}| = 1$, $|\varphi_\mathrm{t}| = 1$, $\zeta \in [0,1]$, and $\eta \in [0,1]$ denote the reflecting and  transmitting amplitudes, respectively.
Since the IOS is a passive device without any active component, we have $\zeta^2 + \eta^2 = 1$ with ignoring the hardware loss \cite{IOS3}, \cite{STAR2}-\cite{STAR5}.

\newcounter{TempEqCnt}
\setcounter{TempEqCnt}{\value{equation}}
\setcounter{equation}{1}
\begin{figure*}[t]
\begin{subequations}
\begin{align}\label{eq:SINR}
\gamma_{\text{r},k_\mathrm{r}} &= \frac{\big|\big({\mathbf{h}^{H}_{\text{r},k_\mathrm{r}}} \bm{\Phi}_{\text{r}} \mathbf{G} + \mathbf{h}_{\text{d},k_\mathrm{r}}^H\big)\mathbf{w}_{\text{r},k_\mathrm{r}}\big|^2}{\big|\big({\mathbf{h}^{H}_{\text{r},k_\mathrm{r}}} \bm{\Phi}_{\text{r}} \mathbf{G} + \mathbf{h}_{\text{d},k_\mathrm{r}}^H\big)(\sum^{j \in \mathcal{K}_\mathrm{r}}_{ j \neq k_\mathrm{r}}\mathbf{w}_{\mathrm{r},j}+\sum_{k_\mathrm{t} \in \mathcal{K}_\mathrm{t}}\mathbf{w}_{\mathrm{t},k_\mathrm{t}})\big|^2+\sigma_\text{r}^2}, \forall k_\mathrm{r} \in \mathcal{K}_\mathrm{r},\\
\gamma_{\text{t},k_\mathrm{t}} &= \frac{\big|{\mathbf{h}^{H}_{\text{t},k_\mathrm{t}}} \bm{\Phi}_{\text{t}} \mathbf{G}\mathbf{w}_{\text{t},k_\mathrm{t}}\big|^2}
{\big|{\mathbf{h}^{H}_{\text{t},k_\mathrm{t}}} \bm{\Phi}_{\text{t}} \mathbf{G}(\sum_{k_\mathrm{r} \in \mathcal{K}_\mathrm{r}}\mathbf{w}_{\mathrm{r},k_\mathrm{r}}+
\sum^{j \in \mathcal{K}_\mathrm{t}}_{ j \neq k_\mathrm{t}}\mathbf{w}_{\mathrm{t},j})\big|^2+\sigma_\text{t}^2},\forall k_\mathrm{t} \in \mathcal{K}_\mathrm{t}.
\end{align}
\end{subequations}
\rule[-0pt]{18.5 cm}{0.05em}\vspace{-0.2 cm}
\end{figure*}
\setcounter{equation}{\value{TempEqCnt}}

\begin{figure}[t]
\centering
 \includegraphics[height= 3 in]{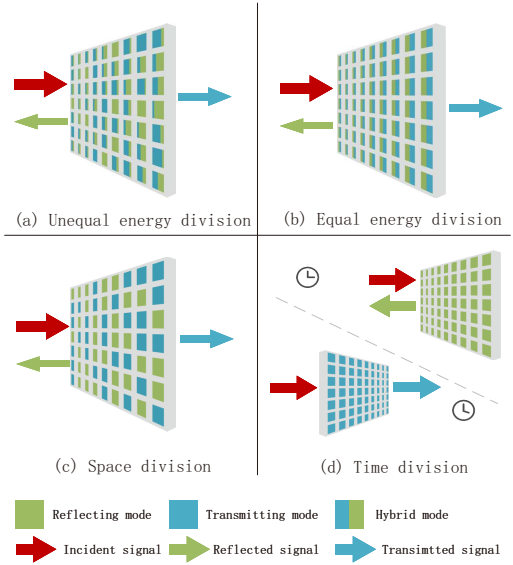}
  \caption{Illustration of different control modes of the IOS.}\label{fig:IOS2}
  \vspace{-0.4 cm}
\end{figure}
\subsection{Different IOS control modes}
The IOS can be operated in different control modes \cite{IOS3}, \cite{STAR}, which can be supported by the state-of-the-art IOS hardware implementation \cite{IOS_hardware_plus}. These modes have different performance and advantages in various communication scenarios, as summarized below.
\begin{enumerate}
  \item Unequal energy division (UED) mode: Each element of the IOS is assumed to simultaneously reflect and transmit with different amplitudes \cite{IOS_Hardware2}, \cite{IOS_Hardware3}, as shown in Fig. 3(a).
    By adjusting the reflecting/transmitting amplitudes and phase-shifts, the incident signals upon each element can be split into transmitted and reflected parts with different amplitudes.
    This mode has the highest DoF and can be applied to various communication scenarios, while it is challenging to design the amplitudes and phase-shifts.
  \item Equal energy division (EED) mode: All elements of the IOS are assumed to simultaneously reflect and transmit with the same amplitude, as shown in Fig. 3(b).
    This common IOS control mode is compatible with the conventional IRS beamforming design algorithm. At the same time, without dynamically adjusting the amplitudes, it suffers performance loss compared with the UED mode.
  \item Space division (SD) mode:
    Partial elements of IOS are in reflection mode while others are in transmission mode, as shown in Fig. 3(c).
    The hardware for SD-mode IOS is easy to implement \cite{STAR}.
    The SD-mode IOS can be deemed as the combination of a reflecting-only and a transmitting-only meta-surfaces with reduced sizes, so it requires a larger number of elements to achieve satisfactory performance.
  \item Time division (TD) mode: All elements of the IOS can only work in reflection mode or transmission mode in each time-slot, as shown in Fig. 3(d).
    Nevertheless, periodically mode change introduces stringent time synchronization requirements, thus increasing the hardware complexity and signaling overhead.
\end{enumerate}
Since the UED mode has the highest DoF and potentially achieves the best performance, we focus on this mode in this paper and develop the joint beamforming design algorithm for UED-based IOS. The algorithms for the other modes can be obtained with the same algorithmic framework with a simple modification.

\subsection{System Model}
\begin{figure}[t]
\centering
  \includegraphics[height= 1.8 in]{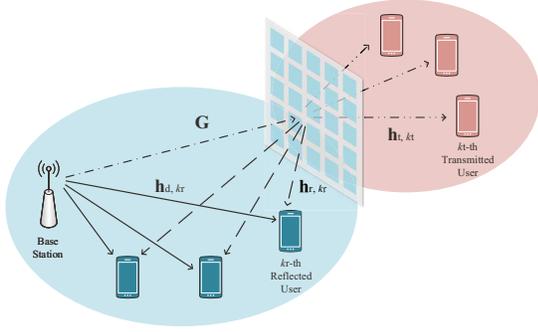}
  \caption{An IOS-assisted multi-user communication system.}\label{fig:System model}
  \vspace{-0.4 cm}
\end{figure}

In this paper, we consider an IOS-assisted downlink multi-user wireless communication system under the UED mode, as shown in Fig. \ref{fig:System model}.
Specifically, a BS equipped with $N_\text{t}$ transmit antennas serves $K_\mathrm{r}$ single-antenna reflected users and $K_\mathrm{t}$ single-antenna transmitted users.
Let $\mathcal{K}_\mathrm{r} \triangleq \{1, \ldots, K_\mathrm{r}\}$
and $\mathcal{K}_\mathrm{t} \triangleq \{1, \ldots, K_\mathrm{t}\}$
denote the sets of reflected users and transmitted users, respectively.
An IOS composed of $M$ elements is deployed to assist the MU-MISO communication.
In the considered communication scenario, the IOS is controlled by an IOS controller through a dedicated control link and only the first-order reflection/transmission is considered due to significant path-loss \cite{IRS_tp1}.
Let
$\mathbf{h}^H_{\text{d},k_\mathrm{r}} \in \mathbb{C}^{1 \times N_\text{t}}$,
$\mathbf{h}^H_{\text{r},k_\mathrm{r}} \in \mathbb{C}^{1 \times M}, \forall k_\mathrm{r} \in \mathcal{K}_\mathrm{r}$, $\mathbf{h}^H_{\text{t},k_\mathrm{t}} \in \mathbb{C}^{1 \times M}, \forall k_\mathrm{t} \in \mathcal{K}_\mathrm{t}$,
and $\mathbf{G} \in \mathbb{C}^{M \times N_\text{t}}$
represent the baseband equivalent channels from the BS to the $k_\mathrm{r}$-th reflected user, from the IOS to the $k_\mathrm{r}$-th reflected user, from the IOS to the $k_\mathrm{t}$-th transmitted user, and from the BS to the IOS, respectively.
In addition, we assume that there is no direct channel for the transmitted users due to blockages.
It is noted that the quasi-static flat-fading Rayleigh channel model is adopted for all channels and we assume that all the channel state information (CSI) is perfectly known at the BS with existing channel estimation approaches \cite{tutorial}, \cite{Channel1}-\cite{Channel6}.

Denote $\mathbf{s}_{\mathrm{r}} \triangleq [s_{\mathrm{r},1}, \ldots, s_{\mathrm{r},K_\mathrm{r}}]^T$
and $\mathbf{s}_{\mathrm{t}} \triangleq [s_{\mathrm{t},1}, \ldots, s_{\mathrm{t},K_\mathrm{t}}]^T$
as the transmitted symbol vectors for the reflected users and the transmitted users, respectively.
At the BS, $\mathbf{s}_{\mathrm{r}}$ and $\mathbf{s}_{\mathrm{t}} $  are respectively precoded by the precoding matrices
$\mathbf{W}_\mathrm{r} \triangleq [\mathbf{w}_{\mathrm{r},1}, \ldots, \mathbf{w}_{\mathrm{r},K_\mathrm{r}}] \in \mathbb{C}^{N_\mathrm{t} \times K_\mathrm{r}}$ and $\mathbf{W}_\mathrm{t}\triangleq [\mathbf{w}_{\mathrm{t},1}, \ldots, \mathbf{w}_{\mathrm{t},K_\mathrm{t}}] \in \mathbb{C}^{N_\mathrm{t} \times K_\mathrm{t}}$.
Hence, the received baseband signals at the $k_\mathrm{r}$-th reflected user and the $k_\mathrm{t}$-th transmitted user can be expressed as
\setcounter{equation}{2}
\begin{equation}
\begin{aligned}
     y_{\mathrm{r},k_\mathrm{r}} &\triangleq  \left(\mathbf{h}^H_{\text{r},k_\mathrm{r}}\bm{\Phi}_{\mathrm{r}}\mathbf{G} + \mathbf{h}^H_{\text{d},k_\mathrm{r}}\right) \left(\mathbf{W}_{\mathrm{r}}\mathbf{s}_{\mathrm{r}}+\mathbf{W}_{\mathrm{t}}\mathbf{s}_{\mathrm{t}} \right)
     + n_{\mathrm{r}}, \forall k_\mathrm{r},\\
     y_{\mathrm{t},k_\mathrm{t}} &\triangleq
     \mathbf{h}^H_{\text{t},k_\mathrm{t}}\bm{\Phi}_{\mathrm{t}}\mathbf{G} \left(\mathbf{W}_{\mathrm{r}}\mathbf{s}_{\mathrm{r}}+\mathbf{W}_{\mathrm{t}}\mathbf{s}_{\mathrm{t}} \right)
     + n_{\mathrm{t}}, \forall k_\mathrm{t},
\end{aligned}
\end{equation}
respectively, where $n_{\mathrm{r}} \sim \mathcal{C}\mathcal{N}(0,\sigma_\text{r}^2) $ and $n_{\mathrm{t}} \sim \mathcal{C}\mathcal{N}(0,\sigma_\text{t}^2)$ denote the additive white Gaussian noise (AWGN) at the reflected users and transmitted users, respectively.
$\bm{\Phi}_{\mathrm{r}} \triangleq \mathrm{diag}\{{\phi}_{\mathrm{r},1}, \ldots, {\phi}_{\mathrm{r},M}\}$
and
$\bm{\Phi}_{\mathrm{t}} \triangleq \mathrm{diag}\{{\phi}_{\mathrm{t},1}, \ldots, {\phi}_{\mathrm{t},M}\}$ are the diagonal reflection and transmit matrices of the IOS, respectively.
Thus, the SINRs of the $k_\mathrm{r}$-th reflected user and the $k_\mathrm{t}$-th transmitted user are respectively given on the top of this page.

\subsection{Transmission Protocol}
\begin{figure}[t]
  \centering
  \vspace{0.2 cm}
  \includegraphics[height= 0.6 in]{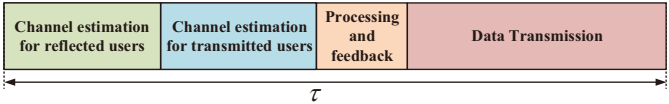}
  \caption{Illustration of the proposed protocol.}\label{fig:protocol}
  \vspace{-0.3 cm}
\end{figure}

In this subsection, we provide a simple transmission protocol for IOS-assisted systems, as shown in Fig. \ref{fig:protocol}.
Since the channel between BS and IOS is usually long-term static, we first estimate the BS-IOS channel off-line based on existing IRS channel estimation methods \cite{Channel1}-\cite{Channel5}.
Then, each channel coherence time is divided into channel estimation phase for reflected users and transmitted users, processing/feedback phase, and a subsequent data transmission phase.
During the channel estimation phase for reflected users, the IOS operates in reflect-only mode (i.e., $\zeta_m = 1, \eta_m = 0, \forall m$). Thus, the BS can only receive the pilot signal from the $K_\text{r}$ users located in the reflective half-space. The BS-IOS-user cascaded channels are first estimated, where the BS-user direct channel can be estimated by proper pilot sequence design \cite{Channel5}. Then the IOS-user channels are estimated by exploiting the off-line estimated BS-IOS channel.
Similarly, in the channel estimation phase for transmitted users, the IOS operates in transmit-only mode (i.e., $\zeta_m = 0, \eta_m = 1, \forall m$) and estimates the channel for transmitted users.
Then, based on the estimated channels, the BS computes the optimal transmit beamforming, IOS reflecting/transmitting phase-shift and amplitude, which are feeded back to the IOS controller.
Finally, data are transmitted to each other using the optimized beamforming.

\section{Algorithm for Power Minimization Problem}
In this section, we aim to jointly optimize the transmit beamformers
$ \mathbf{W}_{\mathrm{r}}$ and $\mathbf{W}_{\mathrm{t}}$,
the phase-shift vectors
$\bm{\varphi}_\mathrm{r} \triangleq [\varphi_\mathrm{r,1}, \ldots, \varphi_{\mathrm{r},M}]$,
and $\bm{\varphi}_\mathrm{t} \triangleq [\varphi_\mathrm{t,1}, \ldots, \varphi_{\mathrm{t},M}]$,
and the IOS reflecting amplitude vector
$\bm{\zeta} \triangleq [\zeta_\mathrm{1}, \ldots, \zeta_M]$,
to minimize the total transmit power for the communication system, subject to the SINR requirement of reflected and transmitted users.
Thus, the power minimization problem can be formulated as
\begin{subequations}
\label{eq:PM problem}
\begin{align}\label{eq:PM problem a}
   \min\limits_{\mathbf{W}_{\mathrm{r}}, \mathbf{W}_{\mathrm{t}},
   \bm{\varphi}_\mathrm{r},\bm{\varphi}_\mathrm{t},\bm{\zeta}}~~&
   \sum_{k_\mathrm{r} \in \mathcal{K}_\mathrm{r}}\left\|\mathbf{w}_{\mathrm{r},k_\mathrm{r}}\right\|^2 +
    \sum_{k_\mathrm{t} \in \mathcal{K}_\mathrm{t}}\left\|\mathbf{w}_{\mathrm{t},k_\mathrm{t}}\right\|^2 \\
   \label{eq:PM problem b}
   \textrm{s.t.}~~& \gamma_{\mathrm{r},k_\mathrm{r}} \geq \Gamma_{\mathrm{r},k_\mathrm{r}}, \forall k_\mathrm{r} \in \mathcal{K}_\mathrm{r},\\
   \label{eq:PM problem c}
    & \gamma_{\mathrm{t},k_\mathrm{t}} \geq \Gamma_{\mathrm{t},k_\mathrm{t}}, \forall k_\mathrm{t} \in \mathcal{K}_\mathrm{t},\\
    \label{eq:PM problem d}
    &\phi_{\mathrm{r},m} = \zeta_m\varphi_{\mathrm{r},m}, \forall m,\\
    \label{eq:PM problem e}
    &\phi_{\mathrm{t},m} = \eta_m \varphi_{\mathrm{t},m}, \forall m,\\
    \label{eq:PM problem f}
    &|\varphi_{\mathrm{r},m}| = 1,\forall m,\\
    \label{eq:PM problem g}
    &|\varphi_{\mathrm{t},m}| = 1,\forall m,\\
    \label{eq:PM problem h}
    & \zeta_m^2 + \eta_m^2 = 1, \forall m,\\
    \label{eq:PM problem i}
    & \zeta_m \in [0,1], \forall m, \\
    \label{eq:PM problem j}
    & \eta_m \in [0,1], \forall m,
\end{align}
\end{subequations}
where $\Gamma_{\mathrm{r},k_\mathrm{r}}$ and $\Gamma_{\mathrm{t},k_\mathrm{t}}$ denote the SINR requirements of the $k_\mathrm{r}$-th reflected user and the $k_\mathrm{t}$-th transmitted user, respectively.
It is noted that the reflecting and transmitting amplitudes are coupled since $\zeta_m^2 + \eta_m^2 = 1$. Therefore, only the reflecting amplitude vector $\bm{\zeta}$ is considered in the optimization (\ref{eq:PM problem}),  and $\bm{\eta}$ can be determined once $\bm{\zeta}$ is designed.
It is challenging to solve the non-convex NP-hard problem (\ref{eq:PM problem}) due to the non-convex constraints (\ref{eq:PM problem b}) and  (\ref{eq:PM problem c}) and the coupled variables $\zeta_m$ and $\eta_m$ in constraints (\ref{eq:PM problem h}).
In order to tackle the difficulties of the non-convex constraints and the correlation between reflection and transmission amplitudes, we develop a three-step algorithm to iteratively solve the original problem.
In particular, we first design the transmit beamforming by utilizing second-order cone programming (SOCP) algorithm.
Then, the reflecting and transmitting phase-shifts $\bm{\varphi}_\mathrm{r}$ and $\bm{\varphi}_\mathrm{t}$ are designed with the manifold optimization. Finally, the IOS reflecting amplitude vector $\bm{\zeta}$ is designed.
In the follows, we will describe the proposed algorithm in details.

\subsection{Transmit Beamforming Design}
When the phase-shift vectors $\bm{\varphi}_\mathrm{r}$, $\bm{\varphi}_\mathrm{t}$,
and the IOS reflecting amplitude vector $\bm{\zeta}$ are fixed,
the sub-problem for optimizing $\mathbf{W}_\mathrm{r}$ and $\mathbf{W}_\mathrm{t}$ is formulated as
\begin{equation}
\label{eq:PM problem2}
\begin{aligned}
   \min\limits_{\mathbf{W}_{\mathrm{r}}, \mathbf{W}_{\mathrm{t}}}~~&
   \sum_{k_\mathrm{r} \in \mathcal{K}_\mathrm{r}}\left\|\mathbf{w}_{\mathrm{r},k_\mathrm{r}}\right\|^2 +
    \sum_{k_\mathrm{t} \in \mathcal{K}_\mathrm{t}}\left\|\mathbf{w}_{\mathrm{t},k_\mathrm{t}}\right\|^2 \\
   \textrm{s.t.}~~& \mathrm{(\ref{eq:PM problem b}) - (\ref{eq:PM problem c})}.
\end{aligned}
\end{equation}
It is obvious that 
this transmit beamforming design problem is a typical power minimization problem and has been well-studied \cite{IRS_tp1}, \cite{IRS_tp2}, which can be transformed into a SOCP problem \cite{SOCP}. Then, it can be easily solved by using the popular convex optimization toolboxes, e.g., CVX \cite{CVX}.

\subsection{IOS Phase Shift Design}
When the transmit beamformers $\mathbf{W}_\mathrm{r}$, $\mathbf{W}_\mathrm{t}$ are obtained and the IOS reflecting amplitude vector $\bm{\zeta}$ is fixed, the problem of designing the IOS beamforming can be expressed as
\begin{equation}
\label{eq:PM problem3}
\begin{aligned}
   \min \limits_{\bm{\varphi}_\mathrm{r},\bm{\varphi}_\mathrm{t}}~~&
   \sum_{k_\mathrm{r} \in \mathcal{K}_\mathrm{r}}\left\|\mathbf{w}_{\mathrm{r},k_\mathrm{r}}\right\|^2 +
    \sum_{k_\mathrm{t} \in \mathcal{K}_\mathrm{t}}\left\|\mathbf{w}_{\mathrm{t},k_\mathrm{t}}\right\|^2 \\
    \textrm{s.t.}~~
    &\mathrm{(\ref{eq:PM problem b}) - (\ref{eq:PM problem g})}.
 \end{aligned}
\end{equation}
Since $\bm{\varphi}_\text{r}$ and $\bm{\varphi}_\text{t}$ do not appear in the objective function, the optimization in fact reduces to a feasibility characterization problem. Note that the feasible domains of $\bm{\varphi}_\text{r}$ and $\bm{\varphi}_\text{t}$ are restricted by different constraints, they can be designed by parallelly solving two problems, respectively. For example, the feasibility-check problem of $\bm{\varphi}_\text{r}$ can be expressed as
\begin{equation}
\label{eq:PM problem_fcheckt}
\begin{aligned}
    \mathrm{Find}~~& \bm{\varphi}_\mathrm{r}\\
    \textrm{s.t.}~~& \gamma_{\mathrm{r},k_\mathrm{r}} \geq \Gamma_{\mathrm{r},k_\mathrm{r}}, \forall k_\mathrm{r} \in \mathcal{K}_\mathrm{r},\\
    &\phi_{\mathrm{r},m} = \zeta_m\varphi_{\mathrm{r},m}, \forall m,\\
    &|\varphi_{\mathrm{r},m}| = 1,\forall m.
\end{aligned}
\end{equation}
\vspace{-0.3 cm}

In the follows, we present the designs of $\bm{\varphi}_\mathrm{r}$ and $\bm{\varphi}_\mathrm{t}$, respectively.
\subsubsection{The Design of $\bm{\varphi}_\mathrm{r}$}
The feasibility-check problem (\ref{eq:PM problem_fcheckt}) has lots of possible solutions. In order to accelerate the convergence, we propose to design $\bm{\varphi}_\mathrm{r}$ with stricter QoS requirement $\gamma_{\mathrm{r},k_\mathrm{r}} \geq \Gamma_{\mathrm{r},k_\mathrm{r}}, \forall k_\mathrm{r} \in \mathcal{K}_\mathrm{r},$ to provide larger DoFs for the power minimization problem in the next iteration. By introducing an auxiliary variable $t$, the new design problem can be formulated as
\begin{equation}
\label{eq:PM problem3t}
\begin{aligned}
   \min \limits_{t, \bm{\varphi}_\mathrm{r}}&~ t\\
   \textrm{s.t.}
   &~~\frac{\Gamma_{\mathrm{r},k_\mathrm{r}}}
   {\gamma_{\mathrm{r},k_\mathrm{r}}} \leq t, \forall k_\mathrm{r},\\
   &~~\phi_{\mathrm{r},m} = \zeta_m\varphi_{\mathrm{r},m}, \forall m,\\
   &~~ |\varphi_{\mathrm{r},m}| = 1,\forall m.
\end{aligned}
\end{equation}
Note that the optimal solution $t^\star$ to (\ref{eq:PM problem3t}) implicitly yields $t^\star \leq 1$ provided $\gamma_{\text{r}, k_\text{r}}$ obtained previously is feasible. The rationale behind solving the problem (\ref{eq:PM problem3t}) lies in providing better QoS via adjusting $\bm{\varphi}_\text{r}$, which brings about a larger margin for the subsequent optimizations of other variables.
Moreover, in order to facilitate the subsequent manifold-based design algorithm, problem (\ref{eq:PM problem3t}) is equivalently transformed into the following min-max problem
\begin{equation}
\label{eq:PM problem4}
\begin{aligned}
   \min \limits_{\bm{\varphi}_\mathrm{r}}&~\max
   \Big\{\frac{\Gamma_{\mathrm{r},1}}{\gamma_{\mathrm{r},1}}, \ldots, \frac{\Gamma_{\mathrm{r},K_\mathrm{r}}}{\gamma_{\mathrm{r},K_\mathrm{r}}}\Big\} \\
   \textrm{s.t.}&~~ \phi_{\mathrm{r},m} = \zeta_m\varphi_{\mathrm{r},m}, \forall m,\\
   &~~ |\varphi_{\mathrm{r},m}| = 1,\forall m.
\end{aligned}
\end{equation}
The problem (\ref{eq:PM problem4}) is a typical fractional programming problem, which can be transformed by the Dinkelbach-type procedure \cite{DT} using an auxiliary variable $\lambda_\text{r} \in \mathbb{R}^+$.
Thus, the objective function of the problem (\ref{eq:PM problem4}) can be reformulated as
\begin{equation}
\small
\label{eq:PM problem4_obj}
\begin{aligned}
&f_{k_\mathrm{r}}(\bm{\varphi}_\mathrm{r},\lambda_\text{r}) =  -(\Gamma_{\mathrm{r},K_\mathrm{r}} +\lambda_\text{r})
\Big|\mathbf{v}^H_{\mathrm{r},k_\mathrm{r},k_\mathrm{r}}\bm{\varphi}_\mathrm{r} + \bar{h}_{\mathrm{r},k_\mathrm{r},k_\mathrm{r}}\Big|^2
+ \Gamma_{\mathrm{r},K_\mathrm{r}}\Big(\sigma_\text{r}^2\\
&+ \sum_{i \in \mathcal{K}_\mathrm{r}} \Big|\mathbf{v}^H_{\mathrm{r},k_\mathrm{r},i}\bm{\varphi}_\mathrm{r} + \bar{h}_{\mathrm{r},k_\mathrm{r},i}\Big|^2
   + \sum_{j \in \mathcal{K}_\mathrm{t}} \Big|\mathbf{v}^H_{\mathrm{t},k_\mathrm{r},j}\bm{\varphi}_\mathrm{r} + \bar{h}_{\mathrm{t},k_\mathrm{r},j}\Big|^2\Big)\\
&= \bm{\varphi}^H_\mathrm{r}\mathbf{B}_{\mathrm{r},k_\mathrm{r}}\bm{\varphi}_\mathrm{r}
        +2\mathfrak{R}\{\bm{\varphi}_\mathrm{r}^H\mathbf{b}_{\mathrm{r},k_\mathrm{r}}\}
        + c_{\mathrm{r},k_\mathrm{r}}, \forall k_\mathrm{r},
\end{aligned}
\end{equation}
where we define the follows
\begin{subequations}
\begin{align}
    \mathbf{v}_{\mathrm{r},k_\mathrm{r},i} &\triangleq [\mathbf{h}^{H}_{\mathrm{r},k_\mathrm{r}}\mathrm{diag}(\bm{\zeta})\mathrm{diag}(\mathbf{G}\mathbf{w}_{\mathrm{r},i})]^H, \forall i \in \mathcal{K}_\mathrm{r}, \forall k_\mathrm{r},\\
    \mathbf{v}_{\mathrm{t},k_\mathrm{r},j} &\triangleq [\mathbf{h}^{H}_{\mathrm{r},k_\mathrm{r}}\mathrm{diag}(\bm{\zeta})\mathrm{diag}(\mathbf{G}\mathbf{w}_{\mathrm{t},j})]^H, \forall j \in \mathcal{K}_\mathrm{t}, \forall k_\mathrm{r},\\
    \bar{h}_{\mathrm{r},k_\mathrm{r},i} & \triangleq \mathbf{h}_{\text{d},k_\mathrm{r}}^{H}\mathbf{w}_{\text{r},i}, \forall i \in \mathcal{K}_\mathrm{r}, \forall k_\mathrm{r} \in \mathcal{K}_\mathrm{r},\\
    \bar{h}_{\mathrm{t},k_\mathrm{r},j} & \triangleq \mathbf{h}_{\text{d},k_\mathrm{r}}^{H}\mathbf{w}_{\text{t},j}, \forall j \in \mathcal{K}_\mathrm{t}, \forall k_\mathrm{r} \in \mathcal{K}_\mathrm{r},
\end{align}
\end{subequations}
for brevity, and $\mathbf{B}_{\mathrm{r},k_\mathrm{r}}$, $\mathbf{b}_{\mathrm{r},k_\mathrm{r}}$, $c_{\mathrm{r},k_\mathrm{r}}$ in quadratic equation can be respectively expressed as
\begin{subequations}
\begin{align}
    \mathbf{B}_{\mathrm{r},k_\mathrm{r}} \triangleq&~
    \Gamma_{\mathrm{r},K_\mathrm{r}}\Big(
    \sum_{i \in \mathcal{K}_\mathrm{r}} \mathbf{v}_{\mathrm{r},k_\mathrm{r},i}\mathbf{v}_{\mathrm{r},k_\mathrm{r},i}^H
    +\sum_{j \in \mathcal{K}_\mathrm{t}} \mathbf{v}_{\mathrm{t},k_\mathrm{r},j}\mathbf{v}_{\mathrm{t},k_\mathrm{r},j}^H \Big)\nonumber\\
    -&(\Gamma_{\mathrm{r},K_\mathrm{r}}+\lambda_\text{r})\mathbf{v}_{\mathrm{r},k_\mathrm{r}, k_\mathrm{r}}\mathbf{v}_{\mathrm{r},k_\mathrm{r},k_\mathrm{r}}^H, \forall k_\mathrm{r} \in \mathcal{K}_\mathrm{r},\\
    \mathbf{b}_{\mathrm{r},k_\mathrm{r}} \triangleq&~
    \Gamma_{\mathrm{r},K_\mathrm{r}}\Big(
    \sum_{i \in \mathcal{K}_\mathrm{r}} \mathbf{v}_{\mathrm{r},k_\mathrm{r},i}\bar{h}_{\mathrm{r},k_\mathrm{r},i}
    +  \sum_{j \in \mathcal{K}_\mathrm{t}} \mathbf{v}_{\mathrm{t},k_\mathrm{r},j}\bar{h}_{\mathrm{t},k_\mathrm{r},j} \Big)\nonumber\\
    -&(\Gamma_{\mathrm{r},K_\mathrm{r}}+\lambda_\text{r})\mathbf{v}_{\mathrm{r},k_\mathrm{r}, k_\mathrm{r}}\bar{h}_{\mathrm{r},k_\mathrm{r}, k_\mathrm{r}}, \forall k_\mathrm{r} \in \mathcal{K}_\mathrm{r},\\
    c_{\mathrm{r},k_\mathrm{r}} \triangleq&~
    \Gamma_{\mathrm{r},K_\mathrm{r}}\Big(
    \sum_{i \in \mathcal{K}_\mathrm{r}}|\bar{h}_{\mathrm{r},k_\mathrm{r},i}|^2 +  \sum_{j \in \mathcal{K}_\mathrm{t}}|\bar{h}_{\mathrm{t},k_\mathrm{r},j}|^2 \Big)\nonumber\\
    -& (\Gamma_{\mathrm{r},K_\mathrm{r}}+\lambda_\text{r})|\bar{h}_{\mathrm{r},k_\mathrm{r},k_\mathrm{r}}|^2
    + \Gamma_{\mathrm{r},K_\mathrm{r}}\sigma_\text{r}^2,\forall k_\mathrm{r}\in \mathcal{K}_\mathrm{r}.
\end{align}
\end{subequations}
Then, the optimal solution to the problem (\ref{eq:PM problem4}) can be obtained by iteratively updating the IOS reflecting phase-shift vector $\bm{\varphi}_\mathrm{r}$ and the auxiliary variable $\lambda_\text{r}$ until convergence.
Firstly, the optimal $\lambda_\text{r}$ can be calculated by
\begin{equation}
\label{eq:lambda}
   \lambda_\text{r}^\star = \max \limits_{k_\mathrm{r} \in \mathcal{K}_\text{r}}\frac{\Gamma_{\text{r},k_\mathrm{r}}}{\gamma_{\text{r},k_\mathrm{r}}}.
\end{equation}
Next, $\bm{\varphi}_\mathrm{r}$ can be updated by solving the following problem
\begin{subequations}
\label{eq:PM problem5}
\begin{align}
 \label{eq:PM problem5a}
\min \limits_{\bm{\varphi}_\mathrm{r}}&\max \limits_{k_\mathrm{r} \in \mathcal{K}_\mathrm{r}}~  f_{k_\mathrm{r}}(\bm{\varphi}_\mathrm{r})\\
 \label{eq:PM problem5b}
 \textrm{s.t.}~&  |\varphi_{\mathrm{r},m}| = 1,\forall m.
\end{align}
\end{subequations}
In order to handle the non-smooth maximizing terms, we attempt to convert the objective (\ref{eq:PM problem5a}) into a continuous and differentiable function.
By exploiting the well-known log-sum-exp method to replace the maximizing function, problem (\ref{eq:PM problem5}) can be transformed into
\begin{subequations}
\label{eq:PM problem6}
\begin{align}
 \label{eq:PM problem6a}
\min \limits_{\bm{\varphi}_\mathrm{r}}&~ \varepsilon\log\sum_{k \in \mathcal{K}_\mathrm{r}}\exp\Big(
    \frac{f_{k_\mathrm{r}}(\bm{\varphi}_\mathrm{r})}{\varepsilon}\Big)\\
 \label{eq:PM problem6b}
 \textrm{s.t.}~&  |\varphi_{\mathrm{r},m}| = 1,\forall m,
\end{align}
\end{subequations}
where $\varepsilon$ is a relatively small positive number to maintain the approximation.
The problem (\ref{eq:PM problem6}) has a differentiable objective with unit modulus, which can be solved by manifold techniques \cite{Liuxing}, \cite{Liuxing2}.
Specifically, the unit modulus constraint (\ref{eq:PM problem6b}) forms an $M$-dimensional complex circle Riemannian manifold, i.e.,
\begin{equation}
\mathcal{M} \triangleq \left\{{\bm{\varphi}_\text{r}} \in \mathbb{C}^{M}:{\varphi}^*_{\text{r},m}{\varphi}_{\text{r},m}=1, \forall m\right\}.
\end{equation}
On the manifold, the direction of movement for each point forms the tangential space, i.e.,
\begin{equation}
T_{\bm{\varphi}_\text{r}}\mathcal{M} \triangleq \left\{\bm{\chi} \in \mathbb{C}^{M}: \mathfrak{R}\left\{\bm{\chi} \odot \bm{\varphi}_\text{r}^{*}\right\}=\mathbf{0}_{M}\right\}.
\end{equation}
Similar to the Euclidean space, the tangent space has a tangent vector in the direction where the objective function decreases fastest, which is called the Riemannian gradient.
Since each point on the manifold has a neighborhood that is isomorphic to the Euclidean space, the gradients of cost functions, distances, angles, etc., have their counterparts on the Riemannian space, and efficient algorithms used on Euclidean space, e.g., the conjugate gradient (CG), are also applicable on the Riemannian manifold \cite{Liuxing3}, \cite{Liuxing4}. Therefore, in the following, we apply the conjugate gradient algorithm on the Riemannian manifold to solve our problem.

To be specific, the problem (\ref{eq:PM problem6}) can be rewritten as
\begin{equation}
\min _{\bm{\varphi}_\text{r} \in \mathcal{M}}
h\left\langle\bm{\varphi}_\text{r}\right\rangle \triangleq
\varepsilon\log\sum_{k \in \mathcal{K}_\mathrm{r}}\exp\Big(
    \frac{f_{k_\mathrm{r}}(\bm{\varphi}_\mathrm{r})}{\varepsilon}\Big),
\end{equation}
which is an unconstrained optimization problem on the Riemannian space $\mathcal{M}$ with the Euclidean gradient of $\left\langle\bm{\varphi}_\text{r}\right\rangle$ as
\begin{equation}\small
\begin{aligned}
\triangledown h\left\langle\bm{\varphi}_\text{r}\right\rangle &\triangleq
 \frac{\varepsilon\sum_{k \in \mathcal{K}_\mathrm{r}}
 \big[\exp(\frac{f_{k_\mathrm{r}}(\bm{\varphi}_\mathrm{r})}{\varepsilon})(\frac{2\mathbf{B}_{\text{r}, k_\text{r}}\bm{\varphi}_\text{r}+2\mathbf{b}_{\text{r}, k_\text{r}}}{\varepsilon})\big]}{\sum_{k \in \mathcal{K}_\mathrm{r}}\exp(\frac{f_{k_\mathrm{r}}(\bm{\varphi}_\mathrm{r})}{\varepsilon})}.
\end{aligned}
\end{equation}
Then, the Riemannian gradient $\operatorname{grad} h\left\langle\bm{\varphi}_\text{r}\right\rangle$ can be obtained by projecting the Euclidean gradient $\triangledown h\left\langle\bm{\varphi}_\text{r}\right\rangle$ onto its corresponding Riemannian tangent space as follows
\begin{equation}
\begin{aligned}
\operatorname{grad} h\left\langle\bm{\varphi}_\text{r}\right\rangle
&=\operatorname{Proj}_{\bm{\varphi}_\text{r}} \nabla h\left\langle\bm{\varphi}_\text{r}\right\rangle\\
&=\nabla h\left\langle\bm{\varphi}_\text{r}\right\rangle-
\Re\left\{\nabla h\left\langle\bm{\varphi}_\text{r}\right\rangle \odot \bm{\varphi}_\text{r}^{*}\right\} \odot \bm{\varphi}_\text{r}.\label{eq:grad}
\end{aligned}
\end{equation}
Then, the search direction $\mathbf{d}_i$ in the $i$-th iteration of the algorithm can be expressed as
\begin{equation}
\mathbf{d}_i \triangleq - \operatorname{grad} h\left\langle\bm{\varphi}_{\text{r},{i-1}}\right\rangle
+ \vartheta_i \mathbf{d}^{\text{t}}_{i-1},\label{eq:di}
\end{equation}
where $\vartheta_i$ is the Polak-Ribiere parameter \cite{Liuxing2}, and $\mathbf{d}^{\text{t}}_{i-1}$ denotes the tangent space of the $\operatorname{grad} h\left\langle\bm{\varphi}_{\text{r},{i-1}}\right\rangle$ \cite{Liuxing4}. After choosing the step size $\omega_i$ using the Armijo backtracking line search method \cite{Liuxing2}, the $i$-th update is given by
\begin{equation}
\bm{\varphi}_{\text{r},{i+1}} = \mathrm{Retr}_{\bm{\varphi}_\text{r}}(\bm{\varphi}_{\text{r},i} + \omega_i\mathbf{d}_i),\label{eq:manifold varphi}
\end{equation}
where $\mathrm{Retr}_{\bm{\varphi}_\text{r}}(\cdot)$ indicates the retraction operation, which maps the points on the tangent space to the manifold. Based on the above analysis, the locally optimal solution to $\bm{\varphi}_\text{r}$ can be obtained using the RCG algorithm as summarized in Algorithm 1.

\begin{algorithm}[t]
\caption{Riemannian conjugate gradient (RCG) algorithm to obtain $\bm{\varphi}_\text{r}$}
\label{alg:SH1}\small
    \begin{algorithmic}[1]
    \REQUIRE $h\left\langle\bm{\varphi}_\text{r}\right\rangle$.
    \ENSURE $\bm{\varphi}^\star_\text{r}$,
    $\mathbf{d}_0 = - \operatorname{grad} h\left\langle\bm{\varphi}_{\text{r},0}\right\rangle$.
    \STATE {Initialize $\bm{\varphi}_{\text{r},0} \in \mathcal{M}$, $\bm{\varphi}_{\text{t}}$, $\bm{\zeta}$.}
    \REPEAT
    \STATE{Choose Polak-Ribiere parameter $\vartheta_i$ \cite{Liuxing2}.}
    \STATE{Calculate search direction $\mathbf{d}_i$ by (\ref{eq:di}).}
    \STATE{Calculate step size $\omega_i$ \cite{Liuxing2}.}
    \STATE{Obtain the update $\bm{\varphi}_{\text{r},{i+1}}$ by (\ref{eq:manifold varphi}).}
    \STATE{Calculate gradient $\operatorname{grad} h\left\langle\bm{\varphi}_\text{r}\right\rangle$ by (\ref{eq:grad}).}
    \UNTIL{\textbf{convergence}}
    \end{algorithmic}
\end{algorithm}

\setcounter{TempEqCnt}{\value{equation}}
\setcounter{equation}{29}
\begin{figure*}[t]
\begin{subequations}
\label{eq:g and h}
\begin{align}
    \bm{\Omega}_{\text{r},p} \triangleq & ~
    \sum_{i \in K_\mathrm{r}}
    \Gamma_{\mathrm{r},p}
    \mathrm{diag}(\bm{\varphi}^H_\mathrm{r}\odot\mathbf{G}\mathbf{w}_{\mathrm{r},i})
    \mathbf{h}_{\mathrm{r},p}\mathbf{h}^H_{\mathrm{r},p}
    \mathrm{diag}(\bm{\varphi}_\mathrm{r}\odot\mathbf{G}\mathbf{w}_{\mathrm{r},i})
    +
    \sum_{i \in K_\mathrm{t}}
    \Gamma_{\mathrm{r},p}
    \mathrm{diag}(\bm{\varphi}^H_\mathrm{r}\odot\mathbf{G}\mathbf{w}_{\mathrm{t},i})
    \mathbf{h}_{\mathrm{r},p}\mathbf{h}^H_{\mathrm{r},p}
    \mathrm{diag}(\bm{\varphi}_\mathrm{r}\odot\mathbf{G}\mathbf{w}_{\mathrm{t},i})\nonumber\\
    & - (1+\kappa)
    \Gamma_{\mathrm{r},p}
    \mathrm{diag}(\bm{\varphi}^H_\mathrm{r}\odot\mathbf{G}\mathbf{w}_{\mathrm{r},p})
    \mathbf{h}_{\mathrm{r},p}\mathbf{h}^H_{\mathrm{r},p}
    \mathrm{diag}(\bm{\varphi}_\mathrm{r}\odot\mathbf{G}\mathbf{w}_{\mathrm{r},p}), \forall p \in K_\mathrm{r},\\
    \bm{\Omega}_{\text{t},p} \triangleq & ~
    \sum_{i \in K_\mathrm{r}}
    \Gamma_{\mathrm{t},p}
    \mathrm{diag}(\bm{\varphi}^H_\mathrm{t}\odot\mathbf{G}\mathbf{w}_{\mathrm{r},i})
    \mathbf{h}_{\mathrm{t},p}\mathbf{h}^H_{\mathrm{t},p}
    \mathrm{diag}(\bm{\varphi}_\mathrm{t}\odot\mathbf{G}\mathbf{w}_{\mathrm{r},i})
    +
    \sum_{i \in K_\mathrm{t}}
    \Gamma_{\mathrm{t},p}
    \mathrm{diag}(\bm{\varphi}^H_\mathrm{t}\odot\mathbf{G}\mathbf{w}_{\mathrm{t},i})
    \mathbf{h}_{\mathrm{t},p}\mathbf{h}^H_{\mathrm{t},p}
    \mathrm{diag}(\bm{\varphi}_\mathrm{t}\odot\mathbf{G}\mathbf{w}_{\mathrm{t},i})\nonumber\\
    & - (1+\kappa)
    \Gamma_{\mathrm{t},p}
    \mathrm{diag}(\bm{\varphi}^H_\mathrm{t}\odot\mathbf{G}\mathbf{w}_{\mathrm{t},p})
    \mathbf{h}_{\mathrm{t},p}\mathbf{h}^H_{\mathrm{t},p}
    \mathrm{diag}(\bm{\varphi}_\mathrm{t}\odot\mathbf{G}\mathbf{w}_{\mathrm{t},p}), \forall p \in K_\mathrm{t},\\
    \mathbf{d}_p \triangleq & ~
    2\Gamma_{\mathrm{r},p}
    \mathfrak{R}\Big\{
    \sum_{q \in K_\mathrm{r}}
    \mathbf{h}^H_{\mathrm{r},p}\mathrm{diag}(\bm{\varphi}^H_\mathrm{r}\odot\mathbf{G}\mathbf{w}_{\mathrm{r},q})
    \mathbf{h}_{\text{d},p}^{H}\mathbf{w}_{\text{r},q}
    +
    \sum_{q \in K_\mathrm{r}}
    \mathbf{h}^H_{\mathrm{r},p}\mathrm{diag}(\bm{\varphi}^H_\mathrm{r}\odot\mathbf{G}\mathbf{w}_{\mathrm{t},q})
    \mathbf{h}_{\text{d},p}^{H}\mathbf{w}_{\text{t},q}\nonumber\\
    & - (1+\kappa)
    \mathbf{h}^H_{\mathrm{r},p}\mathrm{diag}(\bm{\varphi}^H_\mathrm{r}\odot\mathbf{G}\mathbf{w}_{\mathrm{r},q})
    \mathbf{h}_{\text{d},p}^{H}\mathbf{w}_{\text{r},p}
    \Big\}
    , \forall p \in K_\mathrm{r}.
\end{align}
\end{subequations}
\rule[-0pt]{18.5 cm}{0.05em}\vspace{-0.4 cm}
\end{figure*}
\setcounter{equation}{\value{TempEqCnt}}

\subsubsection{The Design of $\bm{\varphi}_\mathrm{t}$}
Similarly, after obtaining the reflecting phase-shift $\bm{\varphi}_\mathrm{t}$, the problem of designing transmitting phase-shift $\bm{\varphi}_\mathrm{t}$ can be expressed as
\begin{equation}
\label{eq:PM problem7}
\begin{aligned}
   \min \limits_{\bm{\varphi}_\mathrm{t}}&~\max
   ~\Big\{\frac{\Gamma_{\mathrm{t},1}}{\gamma_{\mathrm{t},1}}, \ldots, \frac{\Gamma_{\mathrm{t},K_\mathrm{t}}}{\gamma_{\mathrm{t},K_\mathrm{t}}}\Big\} \\
   \textrm{s.t.} &~ |\varphi_{\mathrm{t},m}| = 1,\forall m.
\end{aligned}
\end{equation}
Similarly, each term of the objective function of the problem (\ref{eq:PM problem7}) can be reformulated as
\begin{equation}
f_{k_\mathrm{t}}(\bm{\varphi}_\mathrm{t}, \lambda_\text{t}) = \bm{\varphi}_\mathrm{t}\mathbf{B}_{\mathrm{t},k_\mathrm{t}}\bm{\varphi}_\mathrm{t}
+ \Gamma_{\text{t},k_\mathrm{t}}\sigma_\text{t}^2, \forall k_\mathrm{t},
\end{equation}
with the following parameters
\begin{subequations}
\begin{align}
    \mathbf{b}_{\mathrm{r},k_\mathrm{t},i} \triangleq&~ [\mathbf{h}^{H}_{\mathrm{t},k_\mathrm{t}}\mathrm{diag}(\bm{\eta})\mathrm{diag}(\mathbf{G}\mathbf{w}_{\mathrm{r},i})]^H, \forall i \in \mathcal{K}_\mathrm{r}, \forall k_\mathrm{t},\\
    \mathbf{b}_{\mathrm{t},k_\mathrm{t},j} \triangleq&~  [\mathbf{h}^{H}_{\mathrm{t},k_\mathrm{t}}\mathrm{diag}(\bm{\eta})\mathrm{diag}(\mathbf{G}\mathbf{w}_{\mathrm{t},j})]^H, \forall j \in \mathcal{K}_\mathrm{t}, \forall k_\mathrm{t},\\
    \mathbf{B}_{\mathrm{t},k_\mathrm{t}} \triangleq&~
    \Gamma_{\text{t},k_\mathrm{t}}\Big(
    \sum_{i \in \mathcal{K}_\mathrm{r}} \mathbf{b}_{\mathrm{t},k_\mathrm{t},i}\mathbf{b}_{\mathrm{t},k_\mathrm{t},i}^H
    +\sum_{j \in \mathcal{K}_\mathrm{t}} \mathbf{b}_{\mathrm{t},k_\mathrm{t},j}\mathbf{b}_{\mathrm{t},k_\mathrm{t},j}^H \Big)\nonumber\\
    &-(\Gamma_{\text{t},k_\mathrm{t}}+\lambda_\text{t})\mathbf{b}_{\mathrm{t},k_\mathrm{t}, k_\mathrm{t}}\mathbf{b}_{\mathrm{t},k_\mathrm{t},k_\mathrm{t}}^H, \forall k_\mathrm{t} \in \mathcal{K}_\mathrm{t}.
\end{align}
\end{subequations}
The auxiliary variable $\lambda_\text{t}$ can be iteratively updated by
\begin{equation}
\label{eq:mu}
   \lambda_\text{t}^\star = \max \limits_{k_\mathrm{t} \in \mathcal{K}_\mathrm{t}}\Big\{\frac{\Gamma_{\text{t},k_\mathrm{t}}}{\gamma_{\text{t},k_\mathrm{t}}}\Big\}.
\end{equation}
Then, the transmitting phase-shift $\bm{\varphi}_\mathrm{t}$ is updated by solving the problem
\begin{subequations}\label{eq:PM problem8}
\begin{align}
 \label{eq:PM problem8a}
\min \limits_{\bm{\varphi}_\mathrm{t}}&~ \varepsilon\log\sum_{k_\text{t} \in \mathcal{K}_\mathrm{t}}\exp\Big(
    \frac{-f_{k_\mathrm{t}}(\bm{\varphi}_\mathrm{t})}{\varepsilon}\Big)\\
 \label{eq:PM problem8b}
 \textrm{s.t.}~&  |\varphi_{\mathrm{t},m}| = 1,\forall m,
\end{align}
\end{subequations}
by the manifold optimization.

\subsection{IOS Energy Division Design}
After obtaining transmit beamformers $\mathbf{W}_\mathrm{r}$, $\mathbf{W}_\mathrm{t}$, and the phase-shift vectors $\bm{\varphi}_\mathrm{r}$, $\bm{\varphi}_\mathrm{t}$, the problem of solving the reflecting amplitude vector $\bm{\zeta}$ is also a feasibility check problem.
Similarly, maximizing the minimum weighted SINR is utilized as an alternative objective, which can be expressed as
\begin{subequations}
\label{eq:PM problem9}
\begin{align}
   \max \limits_{\bm{\zeta}}~& \min \big\{\frac{\gamma_{\mathrm{r},1}}{\Gamma_{\mathrm{r},1}}, \ldots, \frac{\gamma_{\mathrm{r},K_\mathrm{r}}}{\Gamma_{\mathrm{r},K_\mathrm{r}}}, \frac{\gamma_{\mathrm{t},1}}{\Gamma_{\mathrm{t},1}}, \ldots, \frac{\gamma_{\mathrm{t},K_\mathrm{t}}}{\Gamma_{\mathrm{t},K_\mathrm{t}}}\big\}\\
    \textrm{s.t.}~~& \zeta^2_m + \eta^2_m =1, \forall m,\\
    ~~& \zeta_m \in [0, 1], \forall m,\\
    ~~& \eta_m \in [0, 1], \forall m.
\end{align}
\end{subequations}
This problem is a non-convex fractional programming problem, and the energy division affects the reflecting and transmitting SINRs in different expressions.
In order to tackle these difficulties, we first utilize the characteristics of the min-max fractional programming to transform the original problem (\ref{eq:PM problem9}) into a quadratic-form problem, which can be expressed as
\begin{equation}
\label{eq:PM problem9t}
\begin{aligned}
   \min \limits_{\bm{\zeta},\kappa}~&\max
   \limits_{p \in \mathcal{K}_\mathrm{r} \atop q \in \mathcal{K}_\mathrm{t}}
   \Big\{\bm{\zeta}^T\bm{\Omega}_{\text{r},p}\bm{\zeta} - 2\mathfrak{R}\{\bm{\zeta}^T\bm{d}_p\},
   \bm{\eta}^T\bm{\Omega}_{\text{t},q}\bm{\eta}
   \Big\}\\
    \textrm{s.t.}~~& \zeta^2_m + \eta^2_m =1, \forall m,\\
    ~~& \zeta_m \in [0, 1], \forall m,\\
    ~~& \eta_m \in [0, 1], \forall m,
\end{aligned}
\end{equation}
where $\bm{\Omega}_{\text{r},p},$ $\bm{\Omega}_{\text{t},q},$ and $\bm{d}_{p}$ are auxiliary variables, which are defined on the top of this page. $\kappa$ is the Dinkelbach auxiliary variable, which is introduced to transform the fractional function into a quadratic-from function.
Then, we can obtain the optimal solution of problem (\ref{eq:PM problem9t}) by iteratively updating $\kappa$ and $\bm{\zeta}$.
With fixed $\bm{\zeta}$, the optimal $\kappa^\star$ can be calculated by a closed-form expression as
\begin{equation}\setcounter{equation}{31}
\label{eq:kappa}
    \kappa^\star  = \max\Big\{
    \frac{\Gamma_{\mathrm{r},1}}{\gamma_{\mathrm{r},1}}, \ldots, \frac{\Gamma_{\mathrm{r},K_\mathrm{r}}}{\gamma_{\mathrm{r},K_\mathrm{r}}}, \frac{\Gamma_{\mathrm{t},1}}{\gamma_{\mathrm{t},1}}, \ldots, \frac{\Gamma_{\mathrm{t},K_\mathrm{t}}}{\gamma_{\mathrm{t},K_\mathrm{t}}}
    \Big\}.
\end{equation}
Next, by exploiting the log-sum-exp method, the objective function of problem (\ref{eq:PM problem9t}) can be transformed into
\begin{equation}
\label{eq:PM problem10}
\begin{aligned}
   g_\text{p}(\bm{\zeta}, \bm{\eta}) = &~ \varepsilon \log \sum_{p \in \mathcal{K}_\mathrm{r}}
   \exp\Big\{
   \frac
   {\bm{\zeta}^T\bm{\Omega}_{\text{r},p}\bm{\zeta} - 2\mathfrak{R}\{\bm{\zeta}^T\bm{d}_p\}}
   {\varepsilon}\Big\} \\
   & ~~+ \varepsilon\sum_{p \in\mathcal{K}_\mathrm{t}}
   \exp\Big\{
   \frac
   {\bm{\eta}^T\bm{\Omega}_{\text{t},p}\bm{\eta}}
   {\varepsilon}\Big\},
\end{aligned}
\end{equation}
where $\varepsilon$ is a relatively small positive number to maintain the approximation.
With the obtained $\kappa$, function (\ref{eq:PM problem10}) is still difficult to be minimized since the reflecting amplitude vector $\bm{\zeta}$ and transmitting amplitude vector $\bm{\eta}$ are embedded into a summation of $K_\text{r} + K_\text{t}$ exponential functions.
To simplify the design, one promising solution is to decompose the joint optimization of the reflecting amplitude vector $\bm{\zeta}$ into $M$ sub-problems, each of which deals with only one variable while fixing others.
The update of $\zeta_m$ is conducted iteratively until the objective value converges.
Towards this end, we first define the following parameters
\begin{subequations}
\begin{align}
\rho_{1,m} \triangleq \varepsilon\log\sum_{p \in\mathcal{K}_\mathrm{t}}
\exp&\Big\{\frac{\bm{\Omega}_{\text{t},p}(m,m)}{\varepsilon}\Big\},\\
\rho_{2,m}  \triangleq \varepsilon\log\sum_{p \in\mathcal{K}_\mathrm{t}}
\exp&\Big\{
   \frac{2\mathfrak{R}\{
   \sum_{m \neq n}\bm{\Omega}_{\text{r},p}(m,n)\eta_n\}}{\varepsilon}\Big\},\\
\rho_{3,m}  \triangleq \varepsilon\log\sum_{p \in\mathcal{K}_\mathrm{r}}
\exp&\Big\{\frac{\bm{\Omega}_{\text{r},p}(m,m)}{\varepsilon}\Big\},\\
\rho_{4,m}  \triangleq \varepsilon\log\sum_{p \in\mathcal{K}_\mathrm{r}}
\exp&\Big\{2\mathfrak{R}\{\sum_{m \neq n}
\bm{\Omega}_{\text{t},p}(m,n)\eta_n \nonumber \\&+ \mathbf{d}_p(m)\}/\varepsilon\Big\},
\end{align}
\end{subequations}
for brevity. Then, the sub-problem with respect to the $m$-th element $\zeta_m$ while fixing other elements can be formulated as
\begin{equation}
\label{eq:PM problem11}
\begin{aligned}
   \min \limits_{\zeta_m}~&\max\Big\{ \eta^2_m\rho_{1,m} + \eta_m\rho_{2,m}, ~\zeta_m^2\rho_{3,m} +  \zeta_m\rho_{4,m}\Big\}\\
    \textrm{s.t.}~~& \zeta^2_m + \eta^2_m =1, \forall m,\\
    ~~& \zeta_m \in [0, 1], \forall m,\\
    ~~& \eta_m \in [0, 1], \forall m,
\end{aligned}
\end{equation}
Notice that both terms in the $\mathrm{max}(\cdot)$ function are the classical quadratic form, whose minimal value points $\omega_{m,1}$ and $\omega_{m,2}$ can be easily obtained by checking the first-order optimality condition.
Then, we can obtain the optimal solution $\zeta^\star_m$ by bisection search in a small interval due to the monotonicity of function.

\begin{algorithm}[t]
\caption{Joint Transmit Beamformer, IOS Phase-Shift, and IOS Energy Split Design for the Power Minimization Problem}
\small
\label{alg:SH2}
    \begin{algorithmic}[1]
    \REQUIRE $\mathbf{h}_{\text{r}}^H$, $\mathbf{h}_{\text{t}}^H$, $\mathbf{G}$, $\mathbf{h}_{\text{d}}^H$, $\Gamma_{r}$, $\Gamma_{t}$, $\sigma_\text{r}^2$, $\sigma_\text{t}^2$.
    \ENSURE $\mathbf{w}_{\text{r}}^\star$, $\mathbf{w}_{\text{t}}^\star$, $\bm{\varphi}_{\text{r}}^\star$, $\bm{\varphi}_{\text{t}}^\star$, $\bm{\zeta}^\star$.
    \STATE {Initialize $\bm{\varphi}_{\text{r}}$, $\bm{\varphi}_{\text{t}}$, $\bm{\zeta}$.}
    \WHILE{no convergence for (\ref{eq:PM problem a})}
        \STATE{Calculate $\mathbf{w}_{\text{r}}^\star$ and $\mathbf{w}_{\text{t}}^\star$ by solving (\ref{eq:PM problem3}) with CVX.}
        \WHILE{no convergence for $\lambda_\text{r}$}
            \STATE{Update $\lambda_\text{r}$ by (\ref{eq:lambda}).}
            \STATE {Obtain $\bm{\varphi}_{\mathrm{r}}^\star$ by solving problem (\ref{eq:PM problem6}) by Algorithm 1.}
        \ENDWHILE
        \WHILE{no convergence for $\lambda_\text{t}$}
            \STATE{Update $\lambda_\text{t}$ by (\ref{eq:mu}).}
             \STATE {Obtain $\bm{\varphi}_{\mathrm{t}}^\star$ by solving problem (\ref{eq:PM problem8}) by Algorithm 1.}
        \ENDWHILE
        \WHILE{no convergence for $\kappa$}
            \STATE{Update $\kappa$ by (\ref{eq:kappa}).}
            \FOR {$m=1:M$}
            \STATE{Update $\zeta_m$ by searching in $[\omega_{m,1}, \omega_{m,2}]$.}
            \ENDFOR
        \ENDWHILE
    \ENDWHILE
    \STATE {Return $\mathbf{w}_{\text{r}}^\star$, $\mathbf{w}_{\text{t}}^\star$, $\bm{\varphi}_{\text{r}}^\star$, $\bm{\varphi}_{\text{t}}^\star$, $\bm{\zeta}^\star$.}
    \end{algorithmic}
\end{algorithm}

\subsection{Summary and Complexity Analysis}
\subsubsection{Summary}
Based on the above derivations, the design for the IOS-assisted MU-MISO system is straightforward and summarized in Algorithm 2.
Given appropriate initializations, the transmit beamformer, IOS phase-shift, and IOS reflecting amplitude are iteratively updated until the convergence is reached.
Since the objective value of the problem (\ref{eq:PM problem}) is positive and non-increasing after each iteration of Algorithm 2, the proposed Algorithm 2 can converge to a local optimum point.
\subsubsection{Complexity Analysis}
In each iteration, updating the transmit beamforming $\mathbf{W}_\text{r}$ and $\mathbf{W}_\text{t}$ by solving an SOCP problem has a complexity of $\mathcal{O}(N_\text{t}^{3}(K_\text{r}+K_\text{t})^{1.5})$;
updating the IOS phase-shift $\bm{\varphi}_\text{r}$ and $\bm{\varphi}_\text{r}$ with Riemannian manifold optimizations has a complexity of $\mathcal{O}((K_\text{r}+K_\text{t}+1)N^2_\text{t}M^{2})$;
updating the IOS reflecting amplitude $\zeta$ has a complexity of $\mathcal{O}((K_\text{r}+K_\text{t}+1)N^2_\text{t}M^{2})$.
Therefore, the total complexity of the proposed algorithm is of order $\mathcal{O}(N_\text{t}^{3}(K_\text{r}+K_\text{t})^{1.5} + 3(K_\text{r}+K_\text{t}+1)N^2_\text{t}M^{2})$.

\section{Algorithms for Sum-rate Maximization Problem}
In this section, we investigate the sum-rate maximization problem in the considered IOS-assisted multi-user communication system.
It is worth noting that the power minimization problem and the sum-rate maximization problem cannot be directly converted from one to another. Therefore, an appropriate optimization algorithm should be developed.

Specifically, our goal is to jointly optimize the transmit beamformers $\mathbf{W}_\mathrm{r}$ and $\mathbf{W}_\mathrm{t}$,
the phase-shift vectors $\bm{\varphi}_\mathrm{r}$ and $\bm{\varphi}_\mathrm{t}$, and the IOS reflecting amplitude vector $\bm{\zeta}$ to maximize the sum-rate,
subject to the transmit power budget of the BS,
the constraints of the reflecting and transmitting phase-shifts,
and the relation between reflecting and transmitting amplitudes.
Therefore, the optimization problem can be formulated as
\begin{subequations}
 \label{eq:sum rate problem}
   \begin{align}\label{eq:sum rate problem a}
   \max\limits_{\mathbf{W}_\mathrm{r}, \mathbf{W}_\mathrm{t} \atop \bm{\varphi}_\mathrm{r},\bm{\varphi}_\mathrm{t},\bm{\zeta}}~~&
   \sum_{k_\mathrm{r}\in\mathcal{K}_\mathrm{r}}\log_2(1+\gamma_{\mathrm{r},k_\mathrm{r}})
   +\sum_{k_\mathrm{t}\in\mathcal{K}_\mathrm{t}}\log_2(1+\gamma_{\mathrm{t},k_\mathrm{t}})\\
   \label{eq:sum rate problem b}
   \textrm{s.t.}~~&\sum_{k_\mathrm{r}\in\mathcal{K}_\mathrm{r}}\left\|\mathbf{w}_{\mathrm{r},k_\mathrm{r}}\right\|^2
                             +\sum_{k_\mathrm{t}\in\mathcal{K}_\mathrm{t}}\left\|\mathbf{w}_{\mathrm{t},k_\mathrm{t}}\right\|^2 \leq P,\\
    \label{eq:sum rate problem  c}
    &\phi_{\mathrm{r},m} = \zeta_{m}\varphi_{\mathrm{r},m}, \forall m,\\
    \label{eq:sum rate problem  d}
    &\phi_{\mathrm{t},m} = \eta_{m}\varphi_{\mathrm{t},m}, \forall m,\\
    \label{eq:sum rate problem  e}
    &|\varphi_{\mathrm{r},m}| = 1,\forall m,\\
    \label{eq:sum rate problem  f}
    &|\varphi_{\mathrm{t},m}| = 1,\forall m,\\
    \label{eq:sum rate problem  g}
    & \zeta_m^2 + \eta_m^2 = 1, \forall m,\\
    \label{eq:sum rate problem  h}
    & \zeta_m \in [0, 1], \forall m,\\
    \label{eq:sum rate problem  i}
    & \eta_m \in [0, 1], \forall m,
\end{align}
\end{subequations}
where $P > 0$ denotes the transmit power budget.
Seeking for the solution for this non-convex NP-hard problem is very difficult, not only due to the complicated objective function (\ref{eq:sum rate problem a}) that contains the fractional terms in $\log({\cdot})$, but also because of the coupled variables $\bm{\varphi}_\mathrm{r}$, $\bm{\varphi}_\mathrm{t}$, and $\bm{\zeta}$ in the objective function, constraints (\ref{eq:sum rate problem c}) and (\ref{eq:sum rate problem d}).
Therefore, in the followings we first employ the WMMSE approach to convert the original optimization problem (\ref{eq:sum rate problem}) into a polynomial problem, and then use the BCD method to transform into a more tractable multi-variable/block optimization and iteratively solve for each variable.

\subsection{Problem Reformulation by WMMSE}
Following the derivations in \cite{WMMSE}, scalars $\nu_{\mathrm{r},k_\mathrm{r}}$ and $\nu_{\mathrm{t},k_\mathrm{t}}$ are applied on the signals of the $k_\mathrm{r}$-th reflected user and $k_\mathrm{t}$-th transmitted user, respectively, to estimate the corresponding transmitted signals $s_{\mathrm{r},k_\mathrm{r}}$ and $s_{\mathrm{t},k_\mathrm{t}}$.
Then, the MSE of the $k_\mathrm{r}$-th reflected user and the $k_\mathrm{t}$-th transmitted user can be respectively calculated as
\begin{equation}\label{eq:MSEr}
\begin{aligned}
   \text{MSE}_{\mathrm{r},k_\mathrm{r}} =~
   &\mathbb{E}\big\{(\nu_{\mathrm{r},k_\mathrm{r}}^* y_{\mathrm{r},k_\mathrm{r}}-s_{\mathrm{r},k_\mathrm{r}} )
                                     (\nu_{\mathrm{r},k_\mathrm{r}}^* y_{\mathrm{r},k_\mathrm{r}}-s_{\mathrm{r},k_\mathrm{r}} )^* \big\}  \\
   =~& \sum_{i \in \mathcal{K}_\mathrm{r}}
   \big|\nu_{\mathrm{r},k_\mathrm{r}}^*
   ({\mathbf{h}^{H}_{\text{r},k_\mathrm{r}}}\bm{\Phi}_\mathrm{r}\mathbf{G}+ \mathbf{h}_{\text{d},k_\mathrm{r}}^H)
   \mathbf{w}_{\mathrm{r},i}\big|^2\\
   & + \sum_{j \in \mathcal{K}_\mathrm{t}}
   \big|\nu_{\mathrm{r},k_\mathrm{r}}^*
   ({\mathbf{h}^{H}_{\text{r},k_\mathrm{r}}}\bm{\Phi}_\mathrm{r}\mathbf{G}+ \mathbf{h}_{\text{d},k_\mathrm{r}}^H)
   \mathbf{w}_{\mathrm{t},j}\big|^2\\
    & -2\mathfrak{R}\big\{\nu_{\mathrm{r},k_\mathrm{r}}^*
   ({\mathbf{h}^{H}_{\text{r},k_\mathrm{r}}}
    \bm{\Phi}_\mathrm{r}
    \mathbf{G}+ \mathbf{h}_{\text{d},k_\mathrm{r}}^H)
    \mathbf{w}_{\mathrm{r},k_\mathrm{r}}\big\}\\
    & +|\nu_{\mathrm{r},k_\mathrm{r}}|^2\sigma_\text{r}^2 + 1, \forall k_\mathrm{r},
\end{aligned}
\end{equation}
\begin{equation}\label{eq:MSEt}
\begin{aligned}
   \text{MSE}_{\mathrm{t},k_\mathrm{t}} =~
   & \mathbb{E}\big\{(\nu_{\mathrm{t},k_\mathrm{t}}^* y_{\mathrm{t},k_\mathrm{t}}-s_{\mathrm{t},k_\mathrm{t}} )
                                     (\nu_{\mathrm{t},k_\mathrm{t}}^* y_{\mathrm{t},k_\mathrm{t}}-s_{\mathrm{t},k_\mathrm{t}} )^* \big\}\\
   =~& \sum_{i \in \mathcal{K}_\mathrm{r}}
   \big|\nu_{\mathrm{t},k_\mathrm{t}}^*
   {\mathbf{h}^{H}_{\text{t},k_\mathrm{t}}}\bm{\Phi}_\mathrm{t}\mathbf{G}
   \mathbf{w}_{\mathrm{r},i}\big|^2\\
   & + \sum_{j \in \mathcal{K}_\mathrm{t}}
   \big|\nu_{\mathrm{t},k_\mathrm{t}}^*
   {\mathbf{h}^{H}_{\text{t},k_\mathrm{t}}}\bm{\Phi}_\mathrm{t}\mathbf{G}
   \mathbf{w}_{\mathrm{t},j}\big|^2 +|\nu_{\mathrm{t},k_\mathrm{t}}|^2\sigma_\text{t}^2 \\
   & -2\mathfrak{R}
   \big\{\nu_{\mathrm{t},k_\mathrm{t}}^*{\mathbf{h}^{H}_{\text{t},k_\mathrm{t}}}\bm{\Phi}_\mathrm{t}\mathbf{G}\mathbf{w}_{\mathrm{t},k_\mathrm{t}}\big\} + 1, \forall k_\mathrm{t}.
\end{aligned}
\end{equation}
By introducing the MSE weights $\mu_{\mathrm{r},k_\mathrm{r}}, \mu_{\mathrm{t},k_\mathrm{t}} \in \mathbb{R}^+, \forall k_\mathrm{r}, k_\mathrm{t}$, the sum-rate maximization problem (\ref{eq:sum rate problem}) now can be equivalently reformulated as
\begin{subequations}
\label{eq:sum rate problem reformualte}
\begin{align}\label{eq:sum rate problem reformualte a}
\min \limits_{\mathbf{W}_\mathrm{r}, \mathbf{W}_\mathrm{t}, \bm{\varphi}_\mathrm{r} \atop \bm{\varphi}_\mathrm{t},\bm{\zeta},\bm{\nu},\bm{\mu}}~ &\sum_{k_\mathrm{r} \in\mathcal{K}_\mathrm{r}}
\left(\mu_{\mathrm{r},k_\mathrm{r}}\text{MSE}_{\mathrm{r},k_\mathrm{r}}-\log_2\mu_{\mathrm{r},k_\mathrm{r}}\right)\nonumber +  \\
&\sum_{k_\mathrm{t} \in\mathcal{K}_\mathrm{t}}
\left(\mu_{\mathrm{t},k_\mathrm{t}}\text{MSE}_{\mathrm{t},k_\mathrm{t}}-\log_2\mu_{\mathrm{t},k_\mathrm{t}}\right)\\
\textrm{s.t.}~~& (\text{\ref{eq:sum rate problem b}})-(\text{\ref{eq:sum rate problem i}}),
\end{align}
\end{subequations}
where $\bm{\mu} \triangleq [\mu_{\mathrm{r},1}, \ldots, \mu_{\mathrm{r},K_\mathrm{r}}, \mu_{\mathrm{t},1}, \ldots, \mu_{\mathrm{t},K_\mathrm{t}}]^H$
and $\bm{\nu} \triangleq [\nu_{\mathrm{r},1}, \ldots, \nu_{\mathrm{r},K_\mathrm{r}}, \nu_{\mathrm{t},1}, \ldots, \nu_{\mathrm{t},K_\mathrm{t}}]^H$ denote the auxiliary vectors.
Since the complicated logarithmic and fractional term in the objective function has been transformed into a polynomial term and a simple $\log_2(\cdot)$ term, the reformulated optimization problem (\ref{eq:sum rate problem reformualte}) is tractable.
Then, the obtained multi-variate optimization problem (\ref{eq:sum rate problem reformualte}) can be solved using the typical BCD method.
The details for updating each variable are presented in the next subsection.

\setcounter{TempEqCnt}{\value{equation}}
\setcounter{equation}{40}
\begin{figure*}[t]
\begin{subequations}\label{eq:EF}
\begin{align}
\mathbf{E}_\text{r} &\triangleq \sum_{k_\text{r}\in\mathcal{K}_\text{r}} \mu_{\text{r},k_\text{r}}|\nu_{\text{r},k_\text{r}}|^2
\Big(\sum_{i\in\mathcal{K}_\text{r}}
\text{diag}\{\mathbf{w}_{\text{r},i}^H\mathbf{G}^H\}
\tilde{\mathbf{h}}_{\text{r},k_\text{r}}
\tilde{\mathbf{h}}_{\text{r},k_\text{r}}^H
\text{diag}\{\mathbf{G}\mathbf{w}_{\text{r},i}\}
+\sum_{j\in\mathcal{K}_\text{t}}
\text{diag}\{\mathbf{w}_{\text{t},j}^H\mathbf{G}^H\}
\tilde{\mathbf{h}}_{\text{r},k_\text{r}}
\tilde{\mathbf{h}}_{\text{r},k_\text{r}}^H
\text{diag}\{\mathbf{G}\mathbf{w}_{\text{t},j}\}\Big),
\forall k_\text{r},\\
\mathbf{E}_\text{t} &\triangleq \sum_{k_\text{t}\in\mathcal{K}_\text{t}} \mu_{\text{t},k_\text{t}}|\nu_{\text{t},k_\text{t}}|^2
\Big(\sum_{i\in\mathcal{K}_\text{r}}
\text{diag}\{\mathbf{w}_{\text{r},i}^H\mathbf{G}^H\}
\tilde{\mathbf{h}}_{\text{t},k_\text{t}}
\tilde{\mathbf{h}}_{\text{t},k_\text{t}}^H
\text{diag}\{\mathbf{G}\mathbf{w}_{\text{r},i}\}
+\sum_{j\in\mathcal{K}_\text{t}}
\text{diag}\{\mathbf{w}_{\text{t},j}^H\mathbf{G}^H\}
\tilde{\mathbf{h}}_{\text{t},k_\text{t}}
\tilde{\mathbf{h}}_{\text{t},k_\text{t}}^H
\text{diag}\{\mathbf{G}\mathbf{w}_{\text{t},j}\}\Big),
\forall k_\text{t},\\
\mathbf{f}_\text{r} &\triangleq \sum_{k_\text{r}\in\mathcal{K}_\text{r}}
\mu_{\text{r},k_\text{r}}
\Big[\nu_{\text{r},k_\text{r}}^*\tilde{\mathbf{h}}_{\text{r},k_\text{r}}^H\text{diag}\{\mathbf{G}\mathbf{w}_{\text{r},k_\text{r}}\}
- |\nu_{\text{r},k_\text{r}}|^2
\tilde{\mathbf{h}}_{\text{r},k_\text{r}}^H
\text{diag}\{\mathbf{G}
(\sum_{i\in\mathcal{K}_\text{r}}
\mathbf{w}_{\text{r},i}\mathbf{w}_{\text{r},i}^H
+\sum_{j\in\mathcal{K}_\text{t}}
\mathbf{w}_{\text{t},j}\mathbf{w}_{\text{t},j}^H
)
\mathbf{h}_{\text{d},k_\text{r}}\}
\Big]
,\forall k_\text{r},\\
\mathbf{f}_\text{t} &\triangleq \sum_{k_\text{t}\in\mathcal{K}_\text{t}}
\nu_{\text{t},k_\text{t}}^*\tilde{\mathbf{h}}_{\text{t},k_\text{t}}^H\text{diag}\{\mathbf{G}\mathbf{w}_{\text{t},k_\text{t}}\}, \forall k_\text{t}.
\end{align}
\end{subequations}
\hrulefill
\end{figure*}
\setcounter{equation}{\value{TempEqCnt}}

\subsection{Block Update by BCD}
\subsubsection{Update $\bm{\nu}$}
With given  the transmit beamformers $\mathbf{W}_\mathrm{r}$ and $\mathbf{W}_\mathrm{t}$, the reflecting and transmitting phase-shift vectors $\bm{\varphi}_\mathrm{r}$ and $\bm{\varphi}_\mathrm{t}$, the IOS reflecting amplitude vector $\zeta$, and the MSE weight vector $\bm{\mu}$, the designs of $\nu_{\text{r}, k_\text{r}}$ and $\nu_{\text{t}, k_\text{t}}$ are independent.
Thus, the optimization problems of designing $\bm{\nu}$ can be expressed as
\begin{subequations}
\begin{align}
    \min_{\nu_{\mathrm{r},k_\mathrm{r}}} &~\nu_{\mathrm{r},k_\mathrm{r}}\text{MSE}_{\mathrm{r},k_\mathrm{r}}, \forall k_\mathrm{r},\\
    \min_{\nu_{\mathrm{t},k_\mathrm{t}}} &~\nu_{\mathrm{t},k_\mathrm{t}}\text{MSE}_{\mathrm{t},k_\mathrm{t}}, \forall k_\mathrm{t},
\end{align}
\end{subequations}
which are unconstrained quadratic convex problems.
Thus, individually setting the first-order derivative with respect to $\nu_{\mathrm{r},k_\mathrm{r}}$ and $\nu_{\mathrm{t},k_\mathrm{t}}$ to zero, the optimal solutions $\nu_{\mathrm{r},k_\mathrm{r}}^\star$ and $\nu_{\mathrm{t},k_\mathrm{t}}^\star$ can be easily calculated as
\begin{subequations}\label{eq:optimal nu}
\begin{align}
    \nu_{\mathrm{r},k_\mathrm{r}}^\star =&~ \frac{({\mathbf{h}^{H}_{\text{r},k_\text{r}}} \bm{\Phi}_\text{r}\mathbf{G} + \mathbf{h}_{\text{d},k_\text{r}}^H) \mathbf{w}_{\text{r},k_\text{r}}}
    {\big|\big|({\mathbf{h}^{H}_{\text{r},k_\mathrm{r}}}\bm{\Phi}_\mathrm{r}\mathbf{G} + \mathbf{h}_{\text{d},k_\mathrm{r}}^H)[\mathbf{W}_\text{r},\mathbf{W}_\text{t}]\big|\big|^2 + \sigma_\text{r}^2}, \forall k_\text{r},\\
    \nu_{\mathrm{t},k_\mathrm{t}}^\star =&~ \frac{{\mathbf{h}^{H}_{\text{t},k_\text{t}}} \bm{\Phi}_\text{t}\mathbf{G}\mathbf{w}_{\text{t},k_\text{t}}}
    {\big|\big|({\mathbf{h}^{H}_{\text{t},k_\mathrm{t}}}\bm{\Phi}_\mathrm{t}\mathbf{G})[\mathbf{W}_\text{r},\mathbf{W}_\text{t}]\big|\big|^2
    + \sigma_\text{t}^2}, \forall k_\text{t}.
\end{align}
\end{subequations}

\subsubsection{Update $\bm{\mu}$}
Fixing other variables, the optimization problems with respect to each independent MSE weight $\mu_{\text{r},k_\text{r}}$ and $\mu_{\text{t},k_\text{t}}$ can be formulated as\setcounter{equation}{41}
\begin{subequations}
\label{eq:S_mu}
\begin{align}
      \min_{\mu_{\text{r},k_\text{r}}}~&~\mu_{\text{r},k_\text{r}}\text{MSE}_{\text{r},k_\text{r}}-\log_2 \mu_{\text{r},k_\text{r}}, \label{eq:S_rhor}\\
      \min_{\mu_{\text{t},k_\text{t}}}~&~\mu_{\text{t},k_\text{t}}\text{MSE}_{\text{t},k_\text{t}}-\log_2 \mu_{\text{t},k_\text{t}}. \label{eq:S_rhot}
\end{align}
\end{subequations}
Similarly, the optimal solution $\mu_{\text{r},k_\text{r}}^\star$ and $\mu_{\text{t},k_\text{t}}^\star$ can be obtained by applying the typical first-order optimality condition as
\begin{subequations}
\label{eq:optimal mu}
\begin{align}
    \mu_{\mathrm{r},k_\mathrm{r}}^\star &= \text{MSE}_{\mathrm{r},k_\mathrm{r}}^{-1} \overset{(a)}{=} 1+\gamma_{\mathrm{r},k_\mathrm{r}},\\
    \mu_{\mathrm{t},k_\mathrm{t}}^\star &= \text{MSE}_{\mathrm{t},k_\mathrm{t}}^{-1} \overset{(a)}{=} 1+\gamma_{\mathrm{t},k_\mathrm{t}},
\end{align}
\end{subequations}
where $(a)$ is derived by substituting the optimal $\nu_{\mathrm{r},k_\mathrm{r}}^\star$ and $\nu_{\mathrm{t},k_\mathrm{t}}^\star$ in (\ref{eq:optimal nu}) into $\text{MSE}_{\text{r},k_\text{r}}$ and $\text{MSE}_{\text{t},k_\text{t}}$, respectively.

\subsubsection{Update $\mathbf{W}$}
After obtaining $\bm{\nu}$, $\bm{\mu}$, $\bm{\varphi}_\mathrm{r}$, $\bm{\varphi}_\mathrm{t}$, and $\bm{\zeta}$, the optimization problem for the design of the BS transmit beamformer $\mathbf{W}_\text{r}$ and the transmit beamformer $\mathbf{W}_\text{t}$ can be formulated as
\begin{subequations}
\label{eq:sum rate problem W}
\begin{align}
    \min \limits_{\mathbf{W}_\text{r}, \mathbf{W}_\text{t}}&
    \sum_{k_\text{r}\in\mathcal{K}_\text{r}}\mu_{\text{r},k_\text{r}}\Big(
    ||\overline{\mathbf{h}}^H_{\text{r},k_\text{r}}[\mathbf{W}_{\text{r}}, \mathbf{W}_{\text{t}}]||^2
    -2\mathfrak{R}\{\overline{\mathbf{h}}^H_{\text{r},k_\text{r}}\mathbf{w}_{\text{r},k_\text{r}}\}\Big)\nonumber\\
    +&
    \sum_{k_\text{t}\in\mathcal{K}_\text{t}}\mu_{\text{t},k_\text{t}}\Big(
    ||\overline{\mathbf{h}}^H_{\text{t},k_\text{t}}[\mathbf{W}_{\text{r}}, \mathbf{W}_{\text{t}}]||^2
    -2\mathfrak{R}\{\overline{\mathbf{h}}^H_{\text{t},k_\text{t}}\mathbf{w}_{\text{t},k_\text{t}}\}\Big)\\
    \label{eq:sum rate problem W b}
    \textrm{s.t.}~&\sum_{k_\mathrm{r}\in\mathcal{K}_\mathrm{r}}\left\|\mathbf{w}_{\mathrm{r},k_\mathrm{r}}\right\|^2
                             +\sum_{k_\mathrm{t}\in\mathcal{K}_\mathrm{t}}\left\|\mathbf{w}_{\mathrm{t},k_\mathrm{t}}\right\|^2 \leq P,
\end{align}
\end{subequations}
where we define $\overline{\mathbf{h}}_{\text{r},k_\text{r}}^H \triangleq\nu_{\text{r},k_\text{r}}^* ({\mathbf{h}^{H}_{\text{r},k_\text{r}}}\bm{\Phi}_\text{r}\mathbf{G} + \mathbf{h}_{\text{d},k_\text{r}}^H)$ and $\overline{\mathbf{h}}_{\text{t},k_\text{t}}^H \triangleq\nu_{\text{t},k_\text{t}}^* {\mathbf{h}^{H}_{\text{t},k_\text{t}}}\bm{\Phi}_\text{t}\mathbf{G}$ for brevity.
Note that (\ref{eq:sum rate problem W}) is a convex optimization problem, which can be readily solved by using standard convex optimization algorithms.
In order to reduce the algorithm complexity, we employ the typical \textit{Lagrange multiplier method} to obtain $\mathbf{W}_\text{r}$ and $\mathbf{W}_\text{t}$ in a closed-form solution.
Applying a Lagrange multiplier $\lambda \geq 0$ to the power constraint (\ref{eq:sum rate problem W b}), the Lagrange function of problem (\ref{eq:sum rate problem W}) can be formulated as
\begin{equation}
\label{eq:sum rate W Lagrange function}
\begin{aligned}
    &\mathcal{L}(\mathbf{W}_\text{r}, \mathbf{W}_\text{t}, \lambda)\\
    &\triangleq\sum_{k_\text{r}\in\mathcal{K}_\text{r}}\mu_{\text{r},k_\text{r}}\Big(
    \big|\big|\overline{\mathbf{h}}^H_{\text{r},k_\text{r}}[\mathbf{W}_{\text{r}}, \mathbf{W}_{\text{t}}]\big|\big|^2 -2\mathfrak{R}\big\{\overline{\mathbf{h}}^H_{\text{r},k_\text{r}}\mathbf{w}_{\text{r},k_\text{r}}\big\}\Big)\\
    &~~~~+
    \sum_{k_\text{t}\in\mathcal{K}_\text{t}}\mu_{\text{t},k_\text{t}}\Big(
    \big|\big|\overline{\mathbf{h}}^H_{\text{t},k_\text{t}}[\mathbf{W}_{\text{r}}, \mathbf{W}_{\text{t}}]\big|\big|^2
    -2\mathfrak{R}\{\overline{\mathbf{h}}^H_{\text{t},k_\text{t}}\mathbf{w}_{\text{t},k_\text{t}}\}\Big)\\
    &~~~~+\lambda\Big(\big|\big|[\mathbf{W}_{\text{r}}, \mathbf{W}_{\text{t}}]\big|\big|_F^2 - P\Big)\nonumber
\end{aligned}
\end{equation}
\begin{equation}
\begin{aligned}
    &=\sum_{k_\text{r}\in\mathcal{K}_\text{r}}\Big[
    \mathbf{w}^H_{\text{r},k_\text{r}}
    \big(\sum_{i\in\mathcal{K}_\text{r}}\mu_{\text{r},i}\overline{\mathbf{h}}_{\text{r},i}\overline{\mathbf{h}}^H_{\text{r},i}
    +\sum_{j\in\mathcal{K}_\text{t}}\mu_{\text{t},j}\overline{\mathbf{h}}_{\text{t},j}\overline{\mathbf{h}}^H_{\text{t},j}\big)
    \mathbf{w}_{\text{r},k_\text{r}}\\
    &~~~~-2\mu_{\text{r},k_\text{r}}\mathfrak{R}\Big\{\overline{\mathbf{h}}_{\text{r},k_\text{r}}\mathbf{w}_{\text{r},k_\text{r}}\Big\}
                +\lambda\mathbf{w}^H_{\text{r},k_\text{r}}\mathbf{w}_{\text{r},k_\text{r}}\Big]\\
    &~~~~+\sum_{k_\text{t}\in\mathcal{K}_\text{t}}\Big[
    \mathbf{w}^H_{\text{t},k_\text{t}}
    \big(\sum_{i\in\mathcal{K}_\text{r}}\mu_{\text{r},i}\overline{\mathbf{h}}_{\text{r},i}\overline{\mathbf{h}}^H_{\text{r},i}
    +\sum_{j\in\mathcal{K}_\text{t}}\mu_{\text{t},j}\overline{\mathbf{h}}_{\text{t},j}\overline{\mathbf{h}}^H_{\text{t},j}\big)
    \mathbf{w}_{\text{t},k_\text{t}}\\
    &~~~~-2\mu_{\text{t},k_\text{t}}\mathfrak{R}\Big\{\overline{\mathbf{h}}_{\text{t},k_\text{t}}\mathbf{w}_{\text{t},k_\text{t}}\Big\}
    + \lambda\mathbf{w}^H_{\text{t},k_\text{t}}\mathbf{w}_{\text{t},k_\text{t}}\Big] - \lambda P.
\end{aligned}
\end{equation}
Then, setting
$\frac{\partial\mathcal{L}(\mathbf{W}_\text{r}, \mathbf{W}_\text{t}, \lambda)}{\mathbf{w}_{\text{r},k_\text{r}}} = \mathbf{0}$, and $\frac{\partial\mathcal{L}(\mathbf{W}_\text{r}, \mathbf{W}_\text{t}, \lambda)}{\mathbf{w}_{\text{t},k_\text{t}}} = \mathbf{0}$,
the optimal beamforming vector $\mathbf{w}_{\text{r},k_\text{r}}^\star$ and $\mathbf{w}_{\text{t},k_\text{t}}^\star$ are given by
\begin{subequations}\label{eq:optimal wsk}
\begin{align}
    \mathbf{w}_{\text{r},k_\text{r}}^\star &=
    \mu_{\text{r},k_\text{r}}\Big(\mathbf{H}_\text{w} + \lambda\mathbf{I}_{N_\text{t}}\Big)^{-1}\overline{\mathbf{h}}_{\text{r},k_\text{r}}, \forall k_\text{r}, \\
    \mathbf{w}_{\text{t},k_\text{t}}^\star &=
    \mu_{\text{t},k_\text{t}}\Big(\mathbf{H}_\text{w}   + \lambda\mathbf{I}_{N_\text{t}}\Big)^{-1}\overline{\mathbf{h}}_{\text{t},k_\text{t}}, \forall k_\text{t},
\end{align}
\end{subequations}
where
$\mathbf{H}_\text{w}  \triangleq \sum_{j\in\mathcal{K}_\text{r}}\mu_{\text{r},j}\overline{\mathbf{h}}_{\text{r},j}\overline{\mathbf{h}}^H_{\text{r},j}
    +\sum_{j\in\mathcal{K}_\text{t}}\mu_{\text{t},j}\overline{\mathbf{h}}_{\text{t},j}\overline{\mathbf{h}}^H_{\text{t},j}$,
and the Lagrange multiplier $\lambda \geq 0$ should be guaranteed to satisfy the complementarity slackness condition of the power constraint (\ref{eq:sum rate problem W b}). Plugging the obtained transmit beamformers $\mathbf{w}_{\text{r},k_\text{r}}^\star$ and $\mathbf{w}_{\text{t},k_\text{t}}^\star$ into the power constraint (\ref{eq:sum rate problem W b}), the optimal solution of $\lambda^\star$ can be easily obtained by one-dimensional search methods.

\subsubsection{Update $\bm{\varphi}_\mathrm{r}$, $\bm{\varphi}_\mathrm{t}$}
Since the IOS reflecting phase-shift $\bm{\varphi}_\mathrm{r}$ and transmitting phase-shift $\bm{\varphi}_\mathrm{t}$ are uncoupled with each other, we can design these two vectors by solving two parallel problems with the fixed variables $\bm{\nu}$, $\bm{\mu}$, $\mathbf{W}_\text{r}$, $\mathbf{W}_\text{t}$, and $\bm{\zeta}$.

To separate reflecting and transmitting amplitudes,
$\tilde{\mathbf{h}}_{\text{r},k_\text{r}} \triangleq \mathbf{h}_{\text{r},k_\text{r}} \odot \bm{\zeta}$ and $\tilde{\mathbf{h}}_{\text{t},k_\text{t}} \triangleq \mathbf{h}_{\text{t},k_\text{t}} \odot \bm{\eta}$
are defined to expressed the IOS reflecting and transmitting equivalent channels, respectively.
Thus, the optimization problem for the reflecting phase-shift $\bm{\varphi}_\mathrm{r}$ design is written by
\begin{subequations}
\label{eq:sum rate thetar}
\begin{align}
\label{eq:sum rate thetarA}
    \min \limits_{\bm{\varphi}_\mathrm{r}}
    &\ \bm{\varphi}_\text{r}^H\mathbf{E}_\text{r}\bm{\varphi}_\text{r} - 2\mathfrak{R}\big\{\bm{\varphi}_\text{r}^H\mathbf{f}_\text{r}\big\}\\
    \label{eq:sum rate thetarB}
    \textrm{s.t.} &~~|\varphi_{\text{r},m}|=1,
\end{align}
\end{subequations}
and the design for transmitting phase-shift $\bm{\varphi}_\mathrm{t}$ can be formulated as
\begin{subequations}
\label{eq:sum rate thetat}
\begin{align}
\label{eq:sum rate thetatA}
    \min \limits_{\bm{\varphi}_\mathrm{t}}
    &\ \bm{\varphi}_\text{t}^H\mathbf{E}_\text{t}\bm{\varphi}_\text{t} - 2\mathfrak{R}\big\{\bm{\varphi}_\text{t}^H\mathbf{f}_\text{t}\big\}\\
    \label{eq:sum rate thetatB}
    \textrm{s.t.} &~~|\varphi_{\text{t},m}|=1,
\end{align}
\end{subequations}
where $\mathbf{E}_\text{r}$, $\mathbf{E}_\text{t}$, $\mathbf{f}_\text{r}$, and $\mathbf{f}_\text{t}$ are defined on the top of this page for brevity.
It is obvious that the main difficulty to tackle problems (\ref{eq:sum rate thetar}) and (\ref{eq:sum rate thetat}) is the non-convex unit modulus constraints (\ref{eq:sum rate thetarB}) and (\ref{eq:sum rate thetatB}).
As discussed in the previous section, these problems can be solved by manifold optimization.
However, in order to improve the efficiency of the algorithm, we propose to iterative design each element in the vectors $\bm{\varphi}_\text{r}$ and $\bm{\varphi}_\text{t}$ until convergence.
Taking the reflecting phase-shift design as an example, the quadratic objective function can be split as
\begin{equation}
\begin{aligned}
& \bm{\varphi}_\text{r}^{H} \mathbf{E}_\text{r} \bm{\varphi}_\text{r}-2 \Re\left\{\bm{\varphi}_\text{r}^{H} \mathbf{f}_\text{r}\right\} \\
=& \sum_{m=1}^{M} \sum_{n=1}^{M} \mathbf{E}_\text{r}(m, n) \varphi_{\text{r},m}^{*} \varphi_{\text{r},n}-2 \Re\left\{\sum_{m=1}^{M} \varphi_{\text{r},m}^{*} \mathbf{f}(m)\right\}.
\end{aligned}
\end{equation}
Then the objective function with respect to the element $\varphi_{\text{r},m}$ can be expressed as
\begin{equation}
\begin{aligned}
f\left(\varphi_{\text{r},m}\right)\triangleq&\sum_{n \neq m} \left(\mathbf{E}_\text{r}(m, n) \varphi_{\text{r},m}^{*} \varphi_{\text{r},n}+\mathbf{E}_\text{r}(n, m) \varphi_{\text{r},n}^{*} \varphi_{\text{r},m}\right) \\
&+\mathbf{E}_\text{r}(m, m)\left|\varphi_{\text{r},m}\right|^{2}-2 \Re\left\{\varphi_{\text{r},m}^{*} \mathbf{f}_\text{r}(m)\right\} \\
=& 2 \Re\{(\sum_{n \neq m} \mathbf{E}_\text{r}(m, n) \varphi_{\text{r},n}-\mathbf{f}_\text{r}(m)) \varphi_{\text{r},m}^{*}\} \\
&+\mathbf{E}_\text{r}(m, m)|\varphi_{\text{r},m}|^{2}, \forall m .
\end{aligned}
\end{equation}
Considering the constant amplitude constraint $|\varphi_{\text{r},m}| = 1$ of each element, $\mathbf{E}_\text{r}(m, m)|\varphi_{\text{r},m}|^{2}$ is a constant term and can be ignored.
Thus, the optimal solution of the $m$-th reflecting phase-shift can be determined by
\begin{equation}
\label{eq:sum rate varphir}
\varphi_{\text{r},m}^{\star}=
\frac{\mathbf{f}_\text{r}(m)-\sum_{n \neq m}\mathbf{E}_\text{r}(m, n)\varphi_{\text{r},n}}
     {|\mathbf{f}_\text{r}(m)-\sum_{n \neq m}\mathbf{E}_\text{r}(m, n)\varphi_{\text{r},n}|}, \forall m.
\end{equation}
In a similar way, the optimal solution of the $m$-th transmitting phase-shift can be obtained by
\begin{equation}
\label{eq:sum rate varphit}
\varphi_{\text{t},m}^{\star}=
\frac{\mathbf{f}_\text{t}(m)-\sum_{n \neq m}\mathbf{E}_\text{t}(m, n)\varphi_{\text{t},n}}
     {|\mathbf{f}_\text{t}(m)-\sum_{n \neq m}\mathbf{E}_\text{t}(m, n)\varphi_{\text{t},n}|}, \forall m.
\end{equation}

\subsubsection{Update $\bm{\zeta}$}
With the fixed $\bm{\nu}$, $\bm{\mu}$, $\mathbf{W}_\text{r}$ and $\mathbf{W}_\text{t}$, and $\bm{\varphi}_\mathrm{r}$ and $\bm{\varphi}_\mathrm{t}$, the optimization problem for IOS reflecting amplitude vector $\bm{\zeta}$ design can be written as
\begin{subequations}
\label{eq:sum rate problem zeta}
\begin{align}
\label{eq:sum rate problem zeta_a}
\min \limits_{\bm{\zeta}}~
&\sum_{k_\mathrm{r} \in\mathcal{K}_\mathrm{r}}\mu_{\mathrm{r},k_\mathrm{r}}\text{MSE}_{\mathrm{r},k_\mathrm{r}}
+ \sum_{k_\mathrm{t} \in\mathcal{K}_\mathrm{t}}\mu_{\mathrm{t},k_\mathrm{t}}\text{MSE}_{\mathrm{t},k_\mathrm{t}}\\
\label{eq:sum rate problem zeta_b}
\textrm{s.t.}~~& \zeta^2_m + \eta^2_m = 1,\\
~~& \zeta_m \in [0, 1], \forall m.
\end{align}
\end{subequations}
In order to facilitate the analysis, we define
$\bm{\Xi}_\mathrm{r} \triangleq \mathrm{diag}(\bm{\varphi}_\mathrm{r})$ and
$\bm{\Xi}_\mathrm{t} \triangleq \mathrm{diag}(\bm{\varphi}_\mathrm{t})$.
Then the objective function (\ref{eq:sum rate problem zeta_a}) can be rearranged as
\begin{equation}
\label{eq:f_zeta}
    \bm{\zeta}^H\mathbf{H}_\mathrm{r}\bm{\zeta}- 2\mathfrak{R}\{\bm{\zeta}^H\bm{\beta}_\mathrm{r}\}
    +\bm{\eta}^H\mathbf{H}_\mathrm{t}\bm{\eta}- 2\mathfrak{R}\{\bm{\eta}^H\bm{\beta}_\mathrm{t}\},
\end{equation}
where
\begin{subequations}
\small
\begin{align}
\mathbf{H}_\mathrm{r} \triangleq& \sum_{k_\mathrm{r}\in\mathcal{K}_\mathrm{r}}\mu_{\mathrm{r},k_\mathrm{r}}
                         |\nu_{\mathrm{r},k_\mathrm{r}}|^2\nonumber\\
                        &(\sum_{p\in\mathcal{K}_\mathrm{r}}
                        \bm{\alpha}_{\text{r},k_\mathrm{r},p}\bm{\alpha}^H_{\text{r},k_\mathrm{r},p}+ \sum_{q\in\mathcal{K}_\mathrm{t}}
                        \bm{\alpha}_{\text{t},k_\mathrm{r},q}\bm{\alpha}^H_{\text{t},k_\mathrm{r},q}),\\
\mathbf{H}_\mathrm{t} \triangleq& \sum_{k_\mathrm{t}\in\mathcal{K}_\mathrm{t}}\mu_{\mathrm{t},k_\mathrm{t}}
                         |\nu_{\mathrm{t},k_\mathrm{t}}|^2\nonumber\\
                        &(\sum_{p\in\mathcal{K}_\mathrm{r}}
                        \bm{\beta}_{\text{r},k_\mathrm{t},p}\bm{\beta}^H_{\text{r},k_\mathrm{t},p,}
                        + \sum_{q\in\mathcal{K}_\mathrm{t}}
                        \bm{\beta}_{\text{t},k_\mathrm{t},q}\bm{\beta}^H_{\text{t},k_\mathrm{t},q}),\\
\bm{\varpi}_\mathrm{r} \triangleq& \sum_{k_\mathrm{r}\in\mathcal{K}_\mathrm{r}}\mu_{\mathrm{r},k_\mathrm{r}}\Big(
                         \nu_{\mathrm{r},k_\mathrm{r}}\bm{\alpha}_{\text{r},k_\mathrm{r},k_\mathrm{r}}\nonumber \\
                         -|&\nu_{\mathrm{r},k_\mathrm{r}}|^2
                         \Big[\sum_{p\in\mathcal{K}_\mathrm{r}}\bm{\alpha}_{\text{r},k_\mathrm{r},p}\bar{h}_{\mathrm{r},k_\mathrm{r},p}
                         +\sum_{q\in\mathcal{K}_\mathrm{t}}\bm{\alpha}_{\text{t},k_\mathrm{r},q}\bar{h}_{\mathrm{t},k_\mathrm{r},q}\Big]\Big),\\
\bm{\varpi}_\mathrm{t} \triangleq& \sum_{k_\mathrm{t}\in\mathcal{K}_\mathrm{t}}\mu_{\mathrm{t},k_\mathrm{t}}
                         \nu_{\mathrm{t},k_\mathrm{t}}\bm{\beta}_{\text{t},k_\mathrm{t},k_\mathrm{t}},
\end{align}
\end{subequations}
with
\begin{subequations}
\begin{align}
\bm{\alpha}_{\text{r},i,j} &\triangleq
[{\mathbf{h}^{H}_{\text{r},i}}
\mathrm{diag}(\bm{\Xi}_\mathrm{r}\mathbf{G}\mathbf{w}_{\mathrm{r},j})]^H,
\forall i, j \in \mathcal{K}_\mathrm{r},\\
\bm{\alpha}_{\text{t},i,j} &\triangleq
[{\mathbf{h}^{H}_{\text{r},i}}
\mathrm{diag}(\bm{\Xi}_\mathrm{r}\mathbf{G}\mathbf{w}_{\mathrm{t},j})]^H,
\forall i \in \mathcal{K}_\mathrm{r}, j \in \mathcal{K}_\mathrm{t},\\
\bm{\beta}_{\text{r},i,j} &\triangleq
[{\mathbf{h}^{H}_{\text{t},i}}
\mathrm{diag}(\bm{\Xi}_\mathrm{t}\mathbf{G}\mathbf{w}_{\mathrm{r},j})]^H,
\forall i \in \mathcal{K}_\mathrm{t}, j \in \mathcal{K}_\mathrm{r},\\
\bm{\beta}_{\text{t},i,j} &\triangleq
[{\mathbf{h}^{H}_{\text{t},i}}
\mathrm{diag}(\bm{\Xi}_\mathrm{t}\mathbf{G}\mathbf{w}_{\mathrm{t},j})]^H,
\forall i,j  \in \mathcal{K}_\mathrm{t}.
\end{align}
\end{subequations}
The rearranged objective function is still difficult to solve since the IOS reflecting amplitude vector $\bm{\zeta}$ and transmitting amplitude vector $\bm{\eta}$ affect the value of the objective function in an opposite trend.
To simplify the design, we propose to decompose the joint optimization of the entire vector $\bm{\zeta}$ into $M$ sub-problems, each of which deals with only one entry of $\bm{\zeta}$ while fixing others.  And we bring constraint (\ref{eq:sum rate problem zeta_b}) into the objective function to decouple the reflecting/transmitting amplitude.
This alternative update of $\bm{\zeta}$ is conducted iteratively until the objective value converges.
Toward this end, we first split the objective function (\ref{eq:f_zeta}) as
\begin{equation}
\begin{aligned}
    &\bm{\zeta}^H\mathbf{H}_\mathrm{r}\bm{\zeta}- 2\mathfrak{R}\{\bm{\zeta}^H\bm{\varpi}_\mathrm{r}\}
    +\bm{\eta}^H\mathbf{H}_\mathrm{t}\bm{\eta}- 2\mathfrak{R}\{\bm{\eta}^H\bm{\varpi}_\mathrm{t}\}\\
    = & \sum_{m \in \mathcal{M}}
    \Big(- 2\mathfrak{R}\big\{\zeta_m\bm{\varpi}_\text{r}(m) + \sqrt{1-\zeta^2_m}\bm{\varpi}_\text{t}(m)\big\}\\
        +&\sum_{n \in \mathcal{M}}
        \big(\mathbf{H}_\text{r}(m,n)\zeta_m\zeta_n+\mathbf{H}_\text{t}(m,n)\sqrt{1-\zeta^2_m}\sqrt{1-\zeta^2_n}\big)
    \Big).
\end{aligned}
\end{equation}
It is noticed that $\mathbf{H}_\text{r} = \mathbf{H}^H_\text{r}, \mathbf{H}_\text{t} = \mathbf{H}^H_\text{t}$, and $\zeta_m \in \mathbb{R}$.
When we only consider the $m$-th element while the other elements are fixed, the related objective function is given by
\begin{equation}
\begin{aligned}
    g_m(\zeta_m) \triangleq& \ \delta_{m,1}\zeta_m + \delta_{m,2}\sqrt{1-\zeta^2_m}\\
    &+ \delta_{m,3}\zeta_m^2 + \delta_{m,4}(1-\zeta_m^2), \forall m,
\end{aligned}
\end{equation}
where $\delta_{m,1} \triangleq 2\mathfrak{R}\{(\sum_{n \neq m}\mathbf{H}_\mathrm{r}(m,n)\zeta_n -\bm{\varpi}_\mathrm{r}(m))\} \in \mathbb{R},$
$\delta_{m,2} \triangleq 2\mathfrak{R}\{\sum_{n \neq m}\mathbf{H}_\mathrm{t}(m,n)\sqrt{1-\zeta^2_n} - \bm{\varpi}_\mathrm{t}(m)\} \in \mathbb{R},$
$\delta_{m,3} \triangleq \mathbf{H}_\mathrm{r}(m,m) \in \mathbb{R},$ and
$\delta_{m,4} \triangleq \mathbf{H}_\mathrm{t}(m,m) \in \mathbb{R},$ are real coefficients.
Then, the $m$-th sub-problem with respect to the $m$-th IOS reflecting amplitude $\zeta_m$ is formulated as
\begin{equation}
\begin{aligned}
\label{eq:sub problem zeta_m}
\min \limits_{\zeta_m}~& g_m(\zeta_m)\\
\textrm{s.t.}~~& \zeta_m \in [0, 1], \forall m,
\end{aligned}
\end{equation}
where the optimal solution $\zeta_m^\star$ can be obtained by applying the typical first-order optimality condition, i.e., find the root of the equation
\begin{equation}
\label{eq:sub problem zeta_md}
    2(\delta_{m,3}-\delta_{m,4})\zeta_m + \delta_{m,1} - \delta_{m,2}\frac{\zeta_m}{\sqrt{1-\zeta^2_m}} = 0,
\end{equation}
which can be determined by using the bisection search method.

\subsection{Summary and Complexity Analysis}
\begin{algorithm}[t]\small
\caption{Joint Transmit Beamformer, IOS Phase-Shift, and IOS Energy Split Design for the for Sum-Rate Maximization Problem}
\label{alg:Algorithm 3}
    \begin{algorithmic}[1]
    \REQUIRE $\mathbf{h}_{\text{r},k_\text{r}}^H$, $\mathbf{h}_{\text{t},k_\text{t}}^H$, $\mathbf{G}_s$, $\mathbf{h}_{\text{d},k_\text{r}}^H$, $P$,
    $\forall k_\text{r}\in\mathcal{K}_\text{r}$,
    $\forall k_\text{t}\in\mathcal{K}_\text{t}$,  $\sigma_\text{r}^2$, $\sigma_\text{t}^2$.
    \ENSURE $\mathbf{W}_\text{r}^\star, \mathbf{W}_\text{t}^\star,$ $\bm{\varphi}_\text{r}^\star, \bm{\varphi}_\text{t}^\star,$
    $\bm{\zeta}^\star$.
    \STATE {Initialize
    $\mathbf{W}_\text{r}, \mathbf{W}_\text{t},$ $\bm{\varphi}_\text{r}, \bm{\varphi}_\text{t},$
    $\bm{\zeta}$.}
    \WHILE {no convergence of the objective (\ref{eq:sum rate problem reformualte a})}
    \STATE {Update $\bm{\nu}$ by (\ref{eq:optimal nu}).}
    \STATE {Update $\bm{\mu}$ by (\ref{eq:optimal mu}).}
    \STATE {Update $\mathbf{W}_{\text{r}}$ and $\mathbf{W}_{\text{t}}$ by (\ref{eq:optimal wsk}).}
    \WHILE {no convergence of the objective (\ref{eq:sum rate thetarA})}
    \FOR {$m=1:M$}
    \STATE{Update $\varphi_{\text{r},m}$ by (\ref{eq:sum rate varphir}).}
    \ENDFOR
    \ENDWHILE
    \WHILE {no convergence of the objective (\ref{eq:sum rate thetatA})}
    \FOR {$m=1:M$}
    \STATE{Update $\varphi_{\text{t},m}$ by (\ref{eq:sum rate varphit}).}
    \ENDFOR
    \ENDWHILE
    \WHILE {no convergence of the objective (\ref{eq:sum rate problem zeta_a})}
    \FOR {$m=1:M$}
    \STATE{Update $\zeta_m$ by determine the root of equation (\ref{eq:sub problem zeta_md}).}
    \ENDFOR
    \ENDWHILE
    \ENDWHILE
    \STATE {Return $\mathbf{W}_\text{r}^\star, \mathbf{W}_\text{t}^\star,$ $\bm{\varphi}_\text{r}^\star, \bm{\varphi}_\text{t}^\star,$
    $\bm{\zeta}^\star$.}
    \end{algorithmic}
\end{algorithm}

\subsubsection{Summary}
Based on the above derivations, the joint transmit beamforming and IOS design for sum-rate maximization problem is straightforward and summarized in Algorithm 3.
The variable $\bm{\nu}$, the MSE weight $\bm{\mu}$, the transmit beamformer $\mathbf{W}_\text{r}$ and $\mathbf{W}_\text{t}$, the IOS phase-shift $\varphi_\text{r}$ and $\varphi_\text{t}$, and IOS reflecting amplitude vector $\bm{\zeta}$ can be iteratively updated until the convergence is met.
Note that the objective value of problem (\ref{eq:sum rate problem reformualte}) is positive and non-increasing after obtaining the closed-form solutions at each iteration of Algorithm 3, thus the proposed Algorithm 3 is guaranteed to converge.

\subsubsection{Complexity Analysis}
In each iteration, updating $\bm{\nu}$ has a complexity of
$\mathcal{O}\{(K_\text{r} + K_\text{t})(K_\text{r} + K_\text{t} + 1)N_\text{t}M^2\}$;
updating $\bm{\mu}$ has a complexity of
$\mathcal{O}\{(K_\text{r} + K_\text{t})^2N_\text{t}M^2\}$;
updating transmit beamformer $\mathbf{W}_\text{r}$ and $\mathbf{W}_\text{t}$ requires about
$\mathcal{O}\{N_\text{t}(K_\text{r} + K_\text{t})(3M^2+N^2_\text{t})\}$ operations;
the order of complexity for updating IOS phase-shifts $\bm{\varphi}_\text{r}$, $\bm{\varphi}_\text{t}$ is about
$\mathcal{O}((K_\text{r} + K_\text{t})^2M^2 + I_1(M-1))$;
the IOS reflecting amplitude $\bm{\zeta}$ has a complexity of
$\mathcal{O}(((K_\text{r} + K_\text{t})^2M^2+ 2I_2(M-1))$, where $I_1$ denotes the
number of iterations for updating the IOS phase-shifts and $I_2$ denotes the number of steps for the bisection search.
Therefore, the total complexity of the proposed algorithm is about
$\mathcal{O}((4N_\text{t} + 3)(K_\text{r} + K_\text{t})^2M^2)$ operations under the assumptions $M \gg N_\text{t}, M \gg K_\text{r} + K_\text{t}$.

\section{Simulation Results}
\begin{figure}[t]
\centering
  \includegraphics[width=3 in]{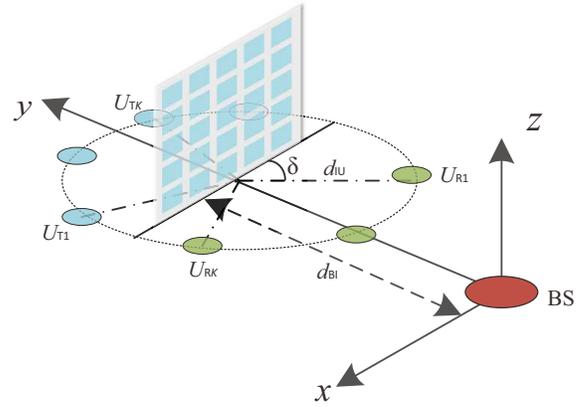}\\
  \caption{An illustration of the relative position among the BS, IOS, and users.}\label{fig:set}
  \vspace{-0.4 cm}
\end{figure}
In this section, extensive simulation results are presented to demonstrate the significance of using the IOS with different control modes in the IOS-assisted multi-user system and the effectiveness of our proposed algorithms.
We assume that the BS equipped with $N_\text{t} = 16$ antennas serves $K_\text{r}$ reflected users and $K_\text{t}$ transmitted users.
An IOS composed of $M = 128$ elements is deployed to assist the downlink communications of the considered system.
The noise power is set as $\sigma_\text{r}^2 = \sigma_\text{t}^2 = -70$ dBm.
The QoS requirements of different users and the power budget of the BS are the same, i.e., $\Gamma = \Gamma_{\text{r},k_\text{r}} = \Gamma_{\text{t},k_\text{t}} = 20 \text{dB}, \forall k_\text{r}, \forall k_\text{t}$,
and $P = 5$ dBW.
In addition, the distance-dependent channel path-loss is modeled as $\eta(d) = C_0(\frac{d}{d_0})^{-\alpha}$, where $C_0 = -30$dB denotes the signal attenuation at the reference distance $d_0 = 1$m, and $\alpha$ denotes the path-loss exponents.
We set the path-loss exponents for the BS-IOS, IOS-user, and BS-user channels as 2.5, 2.8, and 3.5, respectively.

A three-dimensional coordinate system is shown in Fig. \ref{fig:set} to demonstrate the position relationship of different devices in the considered systems.
The BS is located at $(0,0,0)$ and the distance between the IOS and BS is set as $d_{\text{BI}} = 50$m.
Since IOS is deployed to enhance the QoS for both the reflected users and transmitted users,
we assume that all users are randomly distributed at $d_{\text{IU}} = 2$m away from the IOS.
Then the distance between the users and BS can be calculated by $d_{\text{BU}} = \sqrt{d_{\text{BI}}^2 + d_{\text{IU}}^2-2d_{\text{BI}}d_{\text{IU}}\sin\delta}$.
\begin{figure}[t]
\centering
\hspace{-0.6 cm}
  \includegraphics[width = 3.5 in]{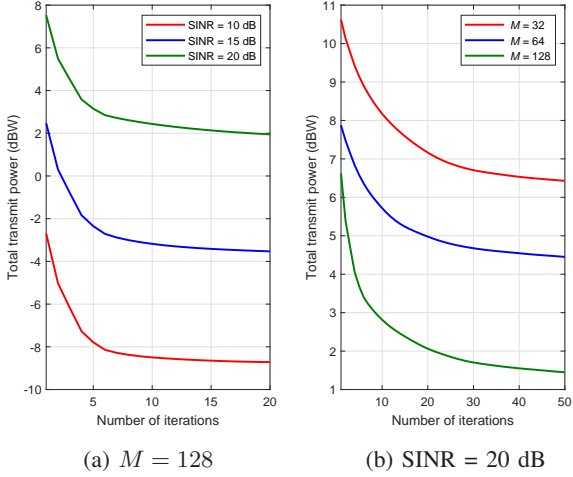}\\
  \small
  \hspace{ 0.1 cm} (a) $M = 128$
  \hspace{ 1.8 cm} (b) SINR = 20 dB
  \vspace{ 0.1 cm}
  \caption{Total transmit power versus the number of iteration ($N_\mathrm{t}= 16$, $K_\text{r}=4$, $K_\text{t}=4$).}
  \label{fig:power_iter}
  \vspace{-0.5 cm}
\end{figure}
\begin{figure}[t]
\centering
\hspace{-0.6 cm}
  \includegraphics[width = 3.5 in]{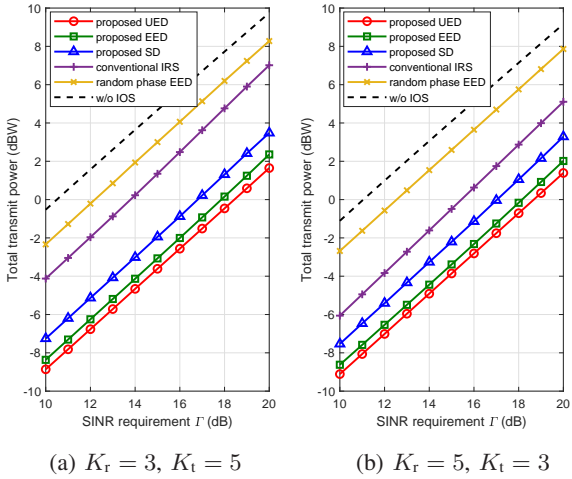}\\
  \small
  \hspace{-0.3 cm} (a) $K_\text{r} = 3$, $K_\text{t} = 5$
  \hspace{ 1.3 cm} (b) $K_\text{r} = 5$, $K_\text{t} = 3$
  \vspace{ 0.1 cm}
  \caption{Total transmit power versus SINR requirement $\Gamma$ ($M = 128, N_\mathrm{t}= 16$).}
  \label{fig:power_sinr}
  \vspace{-0.5 cm}
\end{figure}
\begin{figure}[t]
\centering
\hspace{-0.6 cm}
  \includegraphics[width = 3.5 in]{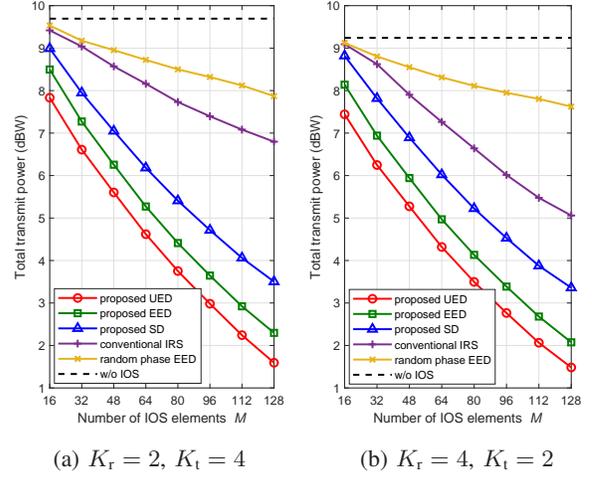}\\
  \small
  \hspace{-0.3 cm} (a) $K_\text{r} = 2$, $K_\text{t} = 4$
  \hspace{ 1.3 cm} (b) $K_\text{r} = 4$, $K_\text{t} = 2$
  \vspace{ 0.1 cm}
  \caption{Total transmit power versus the number of IOS elements $M$ ($\Gamma = 20\mathrm{dB}, N_\mathrm{t} = 16$).}
  \label{fig:power_m}
  \vspace{-0.5 cm}
\end{figure}
\begin{figure}[t]
\centering
\vspace{0.3 cm}
\hspace{-0.2 cm}
  \includegraphics[width = 3.4 in]{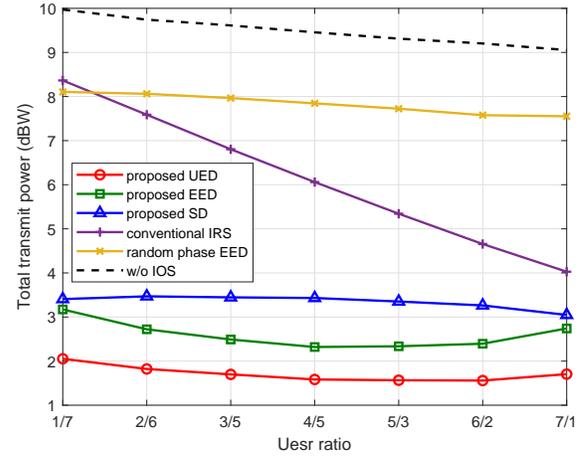}
  \caption{Total transmit power versus the ratio of reflected user to transmitted user ($M = 128, \Gamma = 20\mathrm{dB}, N_\mathrm{t} = 16$).}\label{fig:power ratio}
  \vspace{-0.45 cm}
\end{figure}
\subsection{Power Minimization Problem}
In this subsection, we show the simulation results for the power minimization problem in Figs. \ref{fig:power_iter}-\ref{fig:power ratio}.
The transmit power versus the number of iterations is first presented in Fig. \ref{fig:power_iter} to show the convergence of the proposed algorithm.
It can be observed that the convergence can be achieved within 20 iterations under different settings. These convergence results support the low-complexity implementations.

Next, Fig. \ref{fig:power_sinr} shows the total transmit power as a function of SINR requirement $\Gamma$ under different numbers of reflected/transmitted users.
In the simulation results,  the curves with the legend ``proposed, UED'', ``proposed, EED'', and ``proposed, SD'' show the total transmit power with the proposed algorithm under the UED-mode, EED-mode, and SD-mode, respectively.
It is obvious that ``proposed, UED'' always has the best performance, which verifies the significance of the joint design of transmit beamformers, phase-shifts of IOS, and reflecting/transmitting amplitude of IOS.
The performances of conventional IRS, IOS with random phase-shifts in EED-mode, and without IOS are also presented as benchmarks, which denote as ``conventional IRS", ``random phase EED'', and ``w/o IOS", respectively.
It is noticed that the proposed algorithm with different control modes requires less transmit power than the ``conventional IRS", ``random phase EED'', and ``w/o IOS"  schemes, which validates the advantages of deploying IOS in wireless communication systems.
Moreover, the performance gap between IRS and IOS becomes larger in communication scenarios with more transmitted users. This is because the conventional IRS, which only reflects the signals, cannot effectively serve the users in transmitting half-space.

Fig. \ref{fig:power_m} illustrates the performance of different schemes as a function of the number of IOS elements under different numbers of reflected/transmitted users.
A similar conclusion can be drawn as in Fig. \ref{fig:power_sinr}.
We also observe that the transmit power of all schemes decreases with the increasing number of IOS elements, and the power reduction of our proposed algorithm is more remarkable compared with other competitors.

Moreover, in Fig. \ref{fig:power ratio}, the total transmit power versus the ratio of reflected users to transmitted users is presented. The number on $x$ axis,  for example, $3/5$ represents that the number of reflected users is three and the number of transmitted users is five.
Specifically, the reflected/transmitted users are located at different sides of the IOS, which have different channels and consequently different signal attenuations.
Thus, the communication environment and channel quality significantly change when the user ratio varies.
However, our proposed algorithm with UED mode consistently significantly outperforms the other competitions, and the performance of the UED-mode IOS maintains almost the same with different user ratios.
In addition, since IOS can simultaneously implement reflection and transmission, the performance of IOS is always better than conventional IRS.
Overall, the simulation results verify that the UED-mode IOS with our proposed algorithm can significantly expand the service range and enhance the communication quality in different scenarios.

\begin{figure}[t]
  \centering
  \hspace{-0.6 cm}\includegraphics[width = 3.5 in] {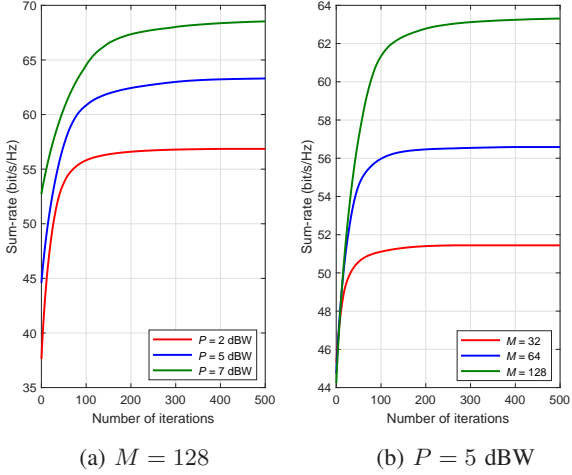}\\
  \small
  \hspace{-0.1 cm} (a) $M = 128$
  \hspace{ 2.0 cm} (b) $P = 5$ dBW
  \vspace{ 0.1 cm}
  \caption{Sum-rate versus the number of iteration ($N_\mathrm{t}= 16$, $K_\text{r}=4$, $K_\text{t}=4$).}\label{fig:sum-rate iter}
  \vspace{-0.5 cm}
\end{figure}
\begin{figure}[t]
  \centering
  \hspace{-0.6 cm}\includegraphics[width = 3.5 in] {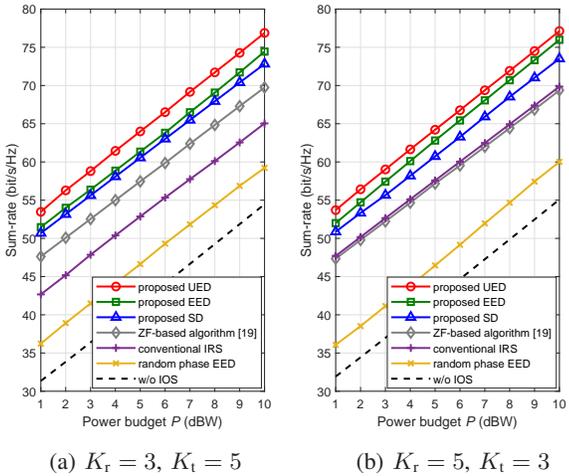}\\
  \small
  \hspace{-0.3 cm} (a) $K_\text{r} = 3$, $K_\text{t} = 5$
  \hspace{ 1.3 cm} (b) $K_\text{r} = 5$, $K_\text{t} = 3$
  \vspace{ 0.1 cm}
  \caption{Sum-rate versus power budget $P$ ($M = 128, N_\mathrm{t}= 16$).}\label{fig:sum-rate p}
  \vspace{-0.45 cm}
\end{figure}
\begin{figure}[t]
  \centering
  \hspace{-0.6 cm}\includegraphics[width = 3.5 in] {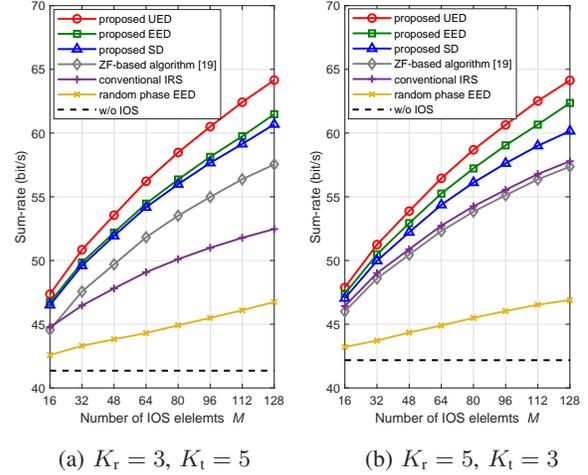}\\
  \small
  \hspace{-0.3 cm} (a) $K_\text{r} = 3$, $K_\text{t} = 5$
  \hspace{ 1.3 cm} (b) $K_\text{r} = 5$, $K_\text{t} = 3$
  \vspace{ 0.1 cm}
  \caption{Sum-rate versus number of IOS elememts $M$ ($P = 5$ dBW, $N_\mathrm{t}= 16$).}\label{fig:sum-rate M}
  \vspace{-0.5 cm}
\end{figure}
\begin{figure}[t]
\centering
  \hspace{-0.2 cm}\includegraphics[width = 3.4 in]{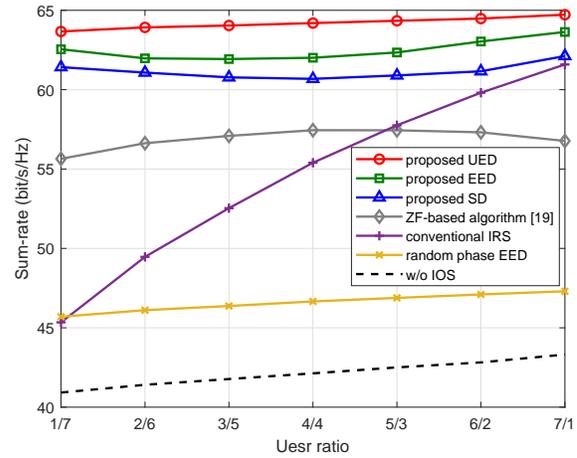}
  \caption{Sum-rate versus the ratio of reflected user to transmitted user ($M = 128, P = 5$ dBW, $N_\mathrm{t} = 16$).}\label{fig:sumrate ratio}
  \vspace{-0.45 cm}
\end{figure}

\subsection{Sum-rate Maximization Problem}
In this subsection, the simulation results for the sum-rate maximization problem are demonstrated in Figs. \ref{fig:sum-rate iter}-\ref{fig:sumrate ratio}.
First of all, similar fast convergence performance as that for the power minimization problem is observed in Fig. \ref{fig:sum-rate iter}.

Fig. \ref{fig:sum-rate p} shows the average sum-rate versus the power budget $P$ under different numbers of reflected/transmitted users.
For better comparison, the curves with the legend ``ZF-based algorithm \cite{IOS1}" show the sum-rate with the ZF-based algorithm \cite{IOS1} using an EED-mode IOS.
It can be seen that more transmit power provides a larger sum-rate for all schemes and the ``proposed UED" always significantly outperforms the others under different settings.
In addition, the performance of our proposed algorithm with different IOS modes is always better than the ZF-based algorithm and conventional IRS, which validates the advantages of our proposed joint design algorithm.

Fig. \ref{fig:sum-rate M} shows the average sum-rate versus the number of IOS elements $M$. The performance gaps between ``proposed, UED'' and others increase when the number of IOS elements increases, which illustrates that the UED-mode IOS has better performance with more elements.
Moreover, the performance of ``ZF-based algorithm \cite{IOS1}" is even worse than the conventional IRS when there are more reflected users and fewer elements, which illustrates the advancement of our proposed algorithm and UED-mode in the IOS-assisted communication systems.

Moreover, the sum-rate versus the ratio of reflected users to transmitted users is presented in Fig. \ref{fig:sumrate ratio}. A similar conclusion as that in Fig. \ref{fig:power ratio} can be obtained.
Interestingly, the performance improvement of ``ZF-based algorithm \cite{IOS1}" is even worse than the conventional IRS when the ratio of reflected user to transmitted user is larger than 5/3. This is because the power ratio is not adjustable in this type of IOS, which is lack of flexibility and leads to energy wastage.
However, our proposed IOS with different modes always has better performance than the conventional IRS, which validates the importance of the design of the IOS energy division.

\section{Conclusions}
In this paper, we investigated the innovative concept of IOS with different control modes, which can provide the dual-functionality of manipulating reflecting and transmitting signals.
Then an IOS-assisted MU-MISO system was considered. The joint transmit beamformers, the reflecting and transmitting phase-shifts of IOS, and the reflecting/tramsitting amplitude designs for both power minimization and sum-rate maximization problems were investigated.
Simulation results verified the advancements of the proposed IOS-assisted wireless communication system and the effectiveness associate joint beamforming design algorithm.
Moreover, there are many issues of IOS-assisted system worth being investigated in future works, including hardware imperfection, fast channel estimation, practical transmission protocol, development of more sophisticated algorithms, as well as learning based methods, etc.

\end{document}